\documentclass[article]{aastex}
\usepackage{emulateapj5}
\lefthead{Zaritsky et al.}
\righthead{SMC Photometric Catalog \& Extinction Map}

\begin{document} 
\title{\bf The Magellanic Clouds Photometric Survey: \break
The Small Magellanic Cloud Stellar Catalog and Extinction Map}

\author{Dennis Zaritsky}
\affil{Steward Observatory, Univ. of Arizona, Tucson, AZ, 85721, email:
dzaritsky@as.arizona.edu} 
\author{Jason Harris}
\affil{Space Telescope Science Institute,
3700 San Martin Dr., Baltimore, MD, 21218 email:jharris@stsci.edu}
\author{Ian B. Thompson}
\affil{Carnegie Observatories, 813 Santa Barbara St., Pasadena, CA 91101,
email:ian@ociw.edu}
\author{Eva K. Grebel}
\affil{Max-Planck-Institut f\"ur Astronomie, K\"onigstuhl 17, D-96177
Heidelberg, Germany, email:grebel@mpia-hd.mpg.de} 
\and 
\author{Philip Massey}
\affil{Lowell Observatory, 1400 West Mars Hill, Flagstaff, AZ, 86001, 
email:massey@lowell.edu}

\begin{abstract}
We present our catalog of $U$, $B$, $V$, and $I$ stellar
photometry
of the central $18 {\rm \ sq.}^\circ$ area of the Small Magellanic Cloud. 
We combine our data with the 2MASS and DENIS 
catalogs to provide, 
when available, $U$ through $K_S$ data for stars. Internal and 
external astrometric and photometric tests using
existing optical photometry ($U$, $B$, and $V$ from Massey's bright star
catalog;
$B, V$, and $I$ from the microlensing
database of OGLE and $I$ from the near-infrared
sky survey DENIS)
are used to determine the observational uncertainties and identify
systematic errors. We fit
stellar atmosphere models to the optical data 
to check the consistency of the photometry for individual stars 
across the passbands and to estimate the line-of-sight
extinction. Finally, we use the estimated line-of-sight extinctions to produce
an extinction map across the Small Magellanic Cloud and investigate
the nature of extinction as a function of stellar population.

\ \ \ \ \ \ \ \ \ \ \ \ \ \ \ \ \hfill\break
\end{abstract} 


\section{Introduction}

The Magellanic Clouds, which are the largest nearby galaxies, provide
our most detailed view of the extragalactic universe. Although their
proximity is generally an advantage, their large angular extent on
the sky has hampered global studies. Historically, stellar catalogs of the 
Clouds have relied
on photographic data (see \cite{hatz89} and \cite{idk90} for some
of the most recent examples). Within the last decade, several
large-scale digital surveys of the Magellanic Clouds have been undertaken.
In the optical bands, the principal ones are the microlensing surveys (MACHO,
\cite{alcock97}; OGLE, \cite{udalski98}; and EROS, \cite{p98}), 
an emission line survey (\cite{smith00}), a bright star survey (\cite{massey01}),
the red optical channel ($I$) of the infrared DENIS survey (\cite{ep97}),
and our Magellanic Clouds Photometric Survey, hereafter MCPS (\cite{zht97}). 

We present the stellar photometric data in catalog form
from the MCPS for the entire Small Magellanic Cloud
(SMC) survey region (roughly $4.5^\circ \times 4^\circ$, where the longer direction
is north-south). 
The principal advantages of these data in comparison to the surveys
listed above are that our data are either deeper, cover a wider
area, or include a larger number of filters (the 
inclusion of $U$ is particularly important
for studies of dust and young stellar populations).
However, as we show in our comparison to these other catalogs, each of the
other surveys has its complementary strengths and we incorporate
data from several of them to augment the MCPS catalog.

In addition to providing the catalog, we construct and analyze extinction maps
of the SMC. As we demonstrated for 
a portion of the LMC (\cite{z99}), the extinction properties in
the Clouds are not only spatially variable, but depend on stellar 
population. Therefore,
for many scientific purposes the catalog alone is insufficient, one
must correct the observed magnitudes and colors for a complex
extinction pattern. 
We describe the MCPS in \S2, discuss detailed comparisons with
previous data to assess the quality of the catalog in
\S3, use the photometry
to generate extinction maps of the SMC for two different stellar
populations in \S4, and present the final catalog in \S5. 

\section{The Data}

The data come from the ongoing Magellanic Cloud Photometric Survey
(\cite{zht97}). Using the Las Campanas Swope telescope (1m) and the
Great Circle Camera (\cite{zsb96}) with a 2K CCD, we obtained drift-scan images 
for both Magellanic Clouds in Johnson $U, B, V, $ and Gunn $I$. The effective
exposure time is between 4 and 5 min for SMC scans and the pixel scale
is 0.7 arcsec pixel$^{-1}$. Typical seeing is $\sim$1.5 arcsec and
scans with seeing worse than $\sim$ 2.5 arcsec are not
accepted. Magnitudes are placed on the Johnson-Kron-Cousins
photometric system (Landolt 1983; 1992).
Data from observing runs from November 1996 to December 1999 are
included in this catalog.  The data are reduced using a pipeline that
utilizes DAOPHOT II (\cite{stet97}) and IRAF\footnote{IRAF is
distributed by the National Optical Observatories, which are operated
by AURA Inc., under contract to the NSF}.  Only stars with both $B$
and $V$ detections are included in the final catalog.

The pipeline for reducing individual scans is a fairly standard
application of DAOPHOT. Each of the 24 scans ($\times$ 4 for the four filters), 
which are either 9500 or 11000
pixels long (depending on whether they are on the eastern or western side
of the SMC survey region) and 2011 pixels wide, 
is divided into 9 by 2 or 11 by 2 subscans that are
roughly 1100 by 1100 pixels, with an overlap of about 100 pixels
between the subscans that enables
us to compare the results from the independent 
photometric reductions. 
There are two sources of discrepancy in the photometry 
from subscans; differing and inconsistent PSF models and aperture corrections,
and nonphotometric conditions. The latter is difficult
to evaluate on a subscan basis because each subscan is 
smaller than the 2K CCD used for the observations, and so adjacent subscans
are not independent of atmospheric variations. Comparing the 
photometry from adjacent scans, rather than subscans, 
enables us to check the observing conditions, which were 
judged by eye during the observing to be photometric almost entirely
over the 4 years. These two classes of
overlaps, subscan and scan, provide complementary internal checks.

The result of the reduction pipeline is a catalog 
of instrumental photometry for each detected star in each filter
and its right ascension and declination. The astrometric solution is 
derived from 
a comparison to stars in the Magellanic Catalogue of Stars (MACS; 
\cite{tu96}), whose coordinates are on the FK5 system.
Solutions are reviewed and iterated if either the number of stars
in the solution is less than 20 over the $\sim$ 12 arcmin $\times 12$ arcmin subscan,
or the rms positional scatter
of the matched stars is larger than 0.5 arcsec. There are only ten
subscans for which we were unable to reduce the positional scatter 
below 0.5 arcsec (and these have rms $<$ 0.6 arcsec), while
the median rms is 0.3 arcsec. 

The instrumental magnitudes of stars in different filters are matched
using a positional match that associates the nearest star on the sky
within an aperture that is 3 times either the positional rms of that
subscan or 1.2 arcsec, whichever is larger. The $V$ frame is used as
the reference and only stars that have a match in the $B$ frame are
retained for the final catalog.  In crowded areas it is possible that
the ``nearest'' star in one filter is not the correct match to the $V$
reference because of the uncertainties in the astrometric solution. 
 We see some evidence of this problem when comparing to
other data and when fitting atmospheric models (stars with highly
anomalous colors), but except near the faint limit of the catalog or in extremely
crowded regions this issue appears
to be a minor problem.  It is evident that one could invest more effort in
attempting to make ``correct'' matches, but the quality of the
photometry in regions where multiple stars are within $\sim$ 1 arcsec
of each other in images with typical 1.5 arcsec seeing is strongly
compromised in any case. In all cases we accept the closest match.
Unlike errors in the photometric calibration,
these errors can be estimated reliably using artificial star
simulations. 

To place each subscan on a self-consistent photometric system, we use the
overlap region (these typically contain several hundred stars in
common in $B$, $V$, and $I$ and several tens in $U$) to measure 
photometric differences. Each subscan has either two (if at the front or
back edge of a scan) or three neighboring subscans within its scan and
one more in the adjacent scan, unless the subscan is at the edge of
the survey region. We calculate the median photometric shift of each
subscan relative to its neighbors and we
find the subscan with the largest offset. The photometry of that
subscan is adjusted by the offset and the process is repeated until
all subscans with a median offset that is greater than 0.02 magnitudes
are corrected. This process converges quickly because of the use of
medians rather than means.  The subscans are then combined to produce
a photometric catalog for each scan. A stellar density map,
constructed from the resulting catalog, is visually inspected
for areas that correspond to scan or subscan regions that are of
anomalously high or low density relative to their neighbors.  The
photometry is adjusted interactively in these cases to correct the
handful of clearly anomalous subscans or scans. This method only addresses
anomalies that are $>$ 0.05 mag. Although we are
concerned that this procedure could lead to systematic drifts from the
correct zeropoints, our photometry is extensively tested by comparison
to external data sets (see \S3.2).

\begin{figure}
\plotone{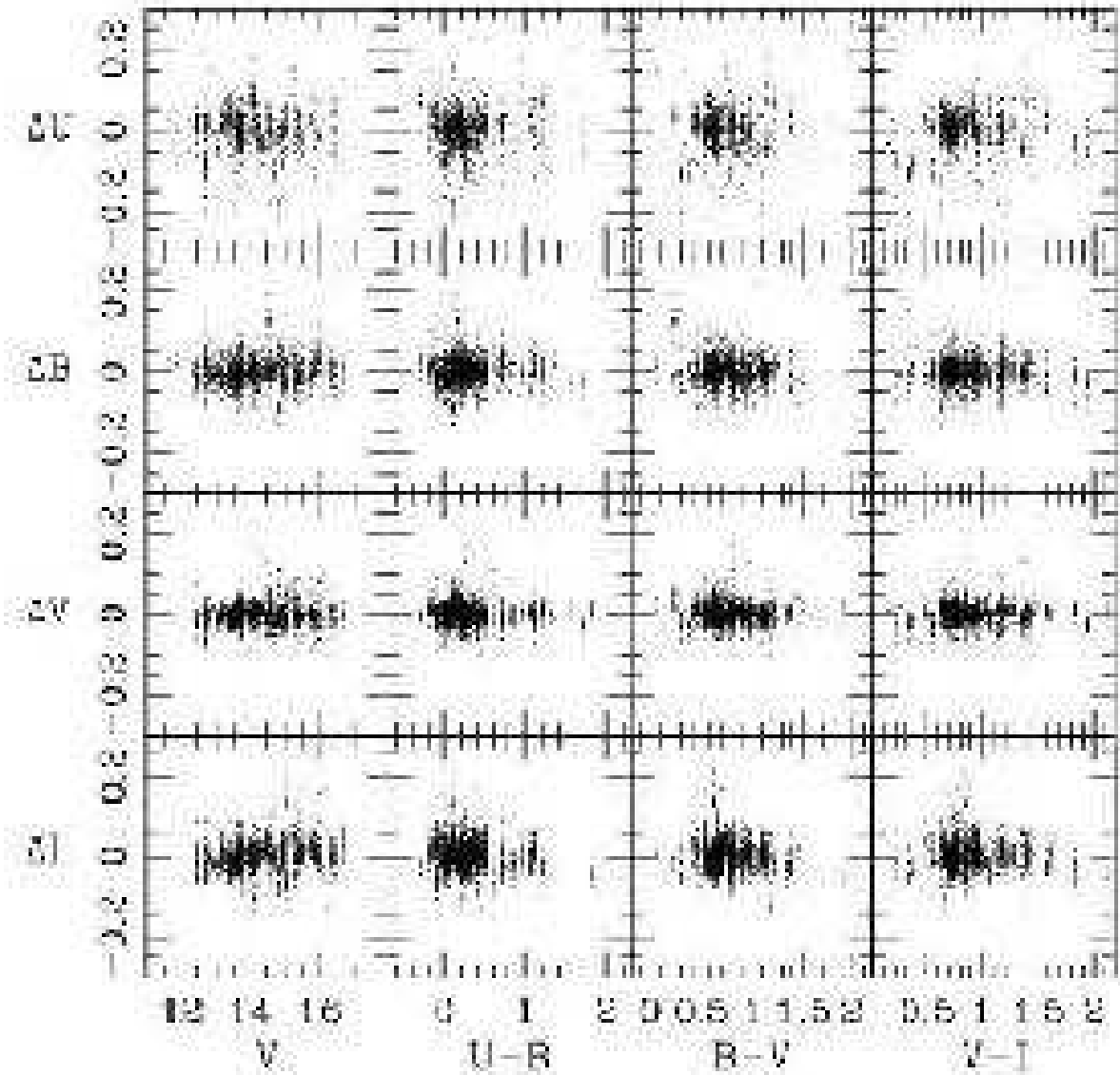}
\caption[stands.ps]{Standard star residuals from the mean photometric
solutions.  Each point represents the measurement for one standard
star sometime during the 4 years of observations. The photometric
solutions are independent for each observing run, but the same
throughout a run.\label{stands} }
\end{figure}

The instrumental magnitudes are placed on the Johnson-Kron-Cousins
Landolt system (1983;
1992). Residuals for the standard stars from our photometric solution
are shown in Figure~\ref{stands} over the four year period of the
observing program.  The photometric solutions involve a zeropoint term, an
airmass term, and one linear color term (using $B-V$ for
$B$ and $V$ and $V-I$ for $I$) or two linear color terms
($U-B$ and $B-V$ separately for $U$).  Over the color range covered by
the standards (see Figure~\ref{stands}) there is little evidence for
additional color terms.  The statistical zeropoint uncertainty is
typically 0.02 mag per run, slightly higher in the $U$ band (0.03 to 0.04 mag)
with only one run having a $U$ band uncertainty as high as 0.047 mag.

\begin{figure}
\figurenum{2}
\vskip -5in
\label{smcimage}
\plotone{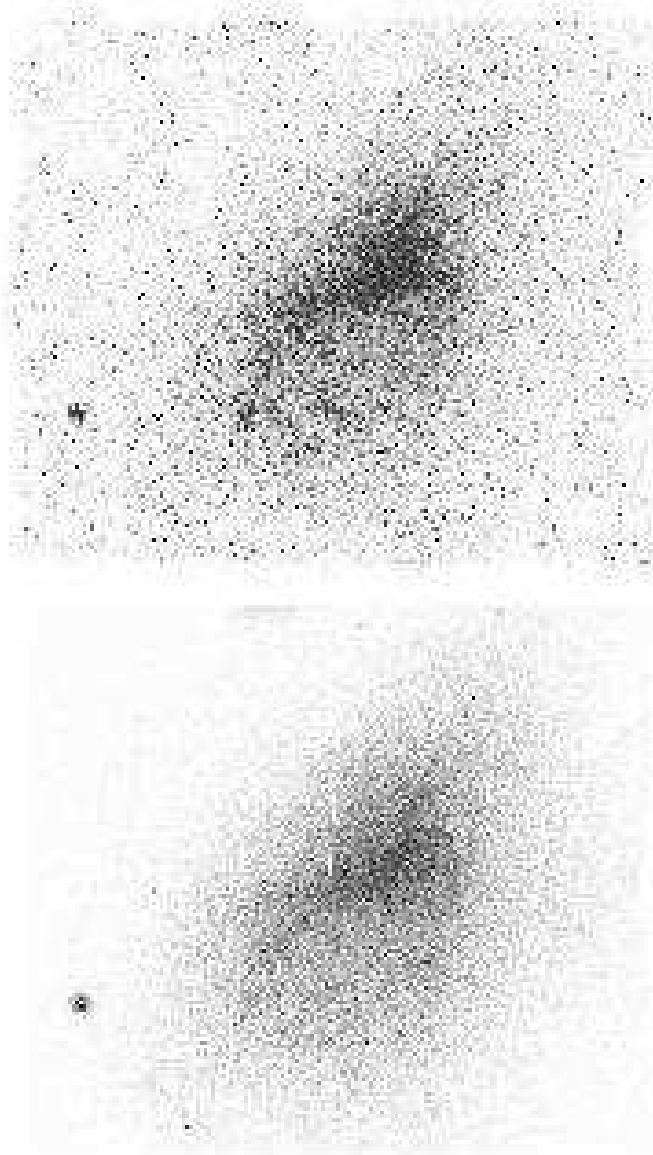}
\figcaption[smcimage.eps]{Grey scale representations of the SMC. The lower
panel is a stellar number density map of stars with $V \le 20$. The upper
panel is a stellar luminosity density map of stars  with $14 < V < 20$. 
North is left and East is to the bottom. The image shows the entire $\sim 4.5^\circ
\times 4^\circ$ survey region. The 
nearly-horizontal white lines are caused by slight gaps between scans, which
occurred if either the $V$ or $B$ scan was slightly misaligned. The large circular
concentration of stars in the northeast is a Galactic globular cluster (NGC 362). The
faint structure at the extreme western edge of the image is 47 Tuc.}
\end{figure}

The catalog of astrometry and photometry for 5,156,057 stars 
is presented as an ASCII table (see Table 1 for a sample).
Columns 1 and 2 contain the right ascension and declination (J2000.0)
for each star. Columns 3-10 contain the pairings of magnitudes and
uncertainties for $U$, $B$, $V$, and $I$ magnitudes.  The subsequent
columns are described in \S5.
$V$ band stellar density and luminosity maps of the SMC are
constructed from the 
catalog using stars with $V < 20$ and shown in Figure~\ref{smcimage}.
The digital catalogs allow one to make analogous images for a variety of populations
(for other examples see the maps of young stars and evolved stars by
\cite{z00}).

\begin{figure}
\vskip -5in
\plotone{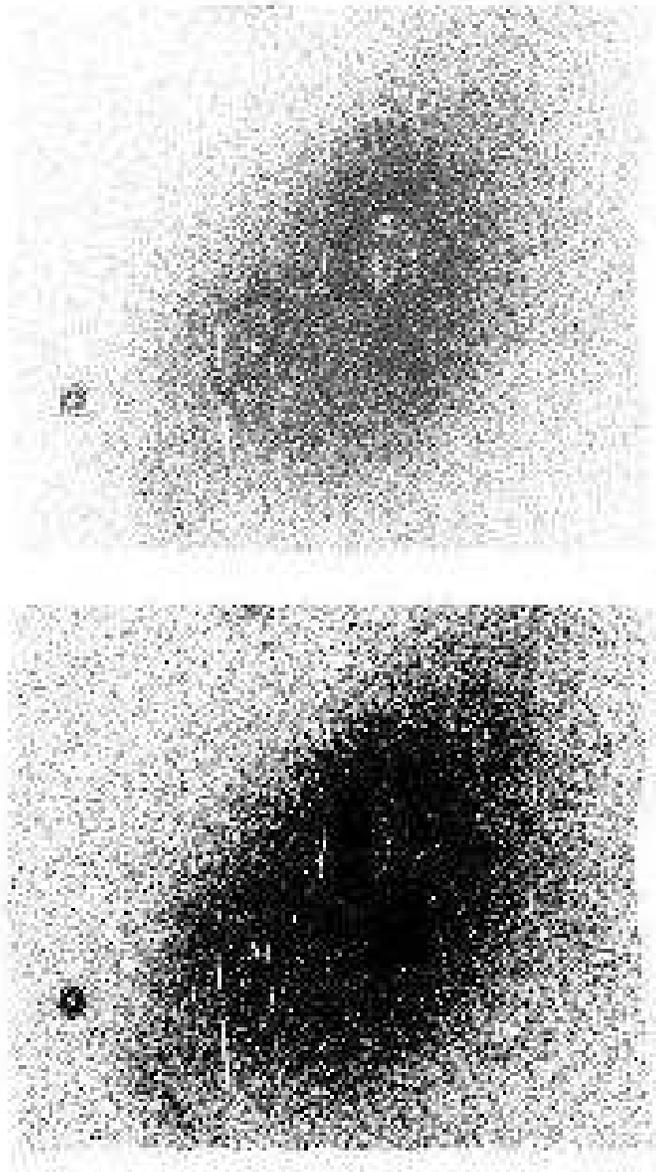}
\figcaption[smc_incomplete.eps]{The stellar density map using stars
with $20 \le V \le 21$. The two panels
show the same map at different contrast levels. 
Incompleteness, due to differing
sensitivity limits caused either by sky brightness,
seeing, or instrumental PSF variations from scan to scan
are revealed by sharp discontinuities in the image. In the upper panel, the
discontinuities between densities in different scans is evident in the 
central four scans, as is the drop in density where the SMC is in
fact most crowded.  At lower density levels (the outer isophotes in the lower
panel) there are no obvious discontinuities at scan edges.}
\label{incomplete}
\end{figure}

The magnitude limit of the survey varies as a function of stellar
crowding. We find little visible evidence for incompleteness for $V <
20$ (Figure~\ref{smcimage}), but the scan edges become visible when
plotting the stellar surface density for stars with $20 < V < 21$
(Figure~\ref{incomplete}). Because of different sky brightness levels,
seeing, and transparency, any completeness variations will be most
noticeable at scan boundaries. We conclude that the stellar density
differences are due to image quality (and the ability to resolve crowded
stars), rather than either to variable sky brightness or 
transparency, because the scans have the larger density
differences in the inner, crowded regions.  In particular, the
densities of the central four scans differ by a factor of 30\% in the
worst central areas for this range of magnitudes. In contrast, the outer isophotes, even at these
magnitudes, are uniform across scan edges (Figure~\ref{incomplete}).
The $U$ and $I$ data are incomplete at brighter magnitudes than the
$B$ and $V$ 
data. The $U$ and $I$
photometry, even in sparse areas, is severely incomplete below $U=21.5$ and
$I = 22$ (comparable limits in the two other
bands are $B=23.5$ and $V= 23$). Any statistical analysis of this catalog
fainter than $V < 20$ requires artificial star tests to determine incompleteness, 
which is becoming significant at these magnitudes.

\section{Testing the Photometry}

\subsection{Internal Tests}

The overlap regions between subscans provide an indication of the
internal reliability of the photometry. Because subscan comparisons
use the same image, they reveal the variation in photometry due to the
different point-spread function (PSF) models and aperture corrections
used for the analysis of each subscan. Although the true PSF of the overlap
stars is obviously the same in both subscans (since they come from the same 
original scan) the PSF models can vary among
subscans due both to atmospheric effects and mechanical effects
(for example, small misalignments or position errors that are
corrected in subsequent moves of the camera) that are not the same
over the entire subscan for each of the overlapping subscans. 
Partially, the initial
motivation for dividing the scans into 1K subscans was that PSF
variations appear manageable across individual subscans of this
size. Although variations are visible even in these smaller regions,
the DAOPHOT PSF is allowed to vary across the image, and the residuals
after subtracting the fitted stars was deemed acceptable. With our
internal and external tests we will determine the extent to which
this conclusion is correct.

\begin{figure}
\figurenum{4}
\plotone{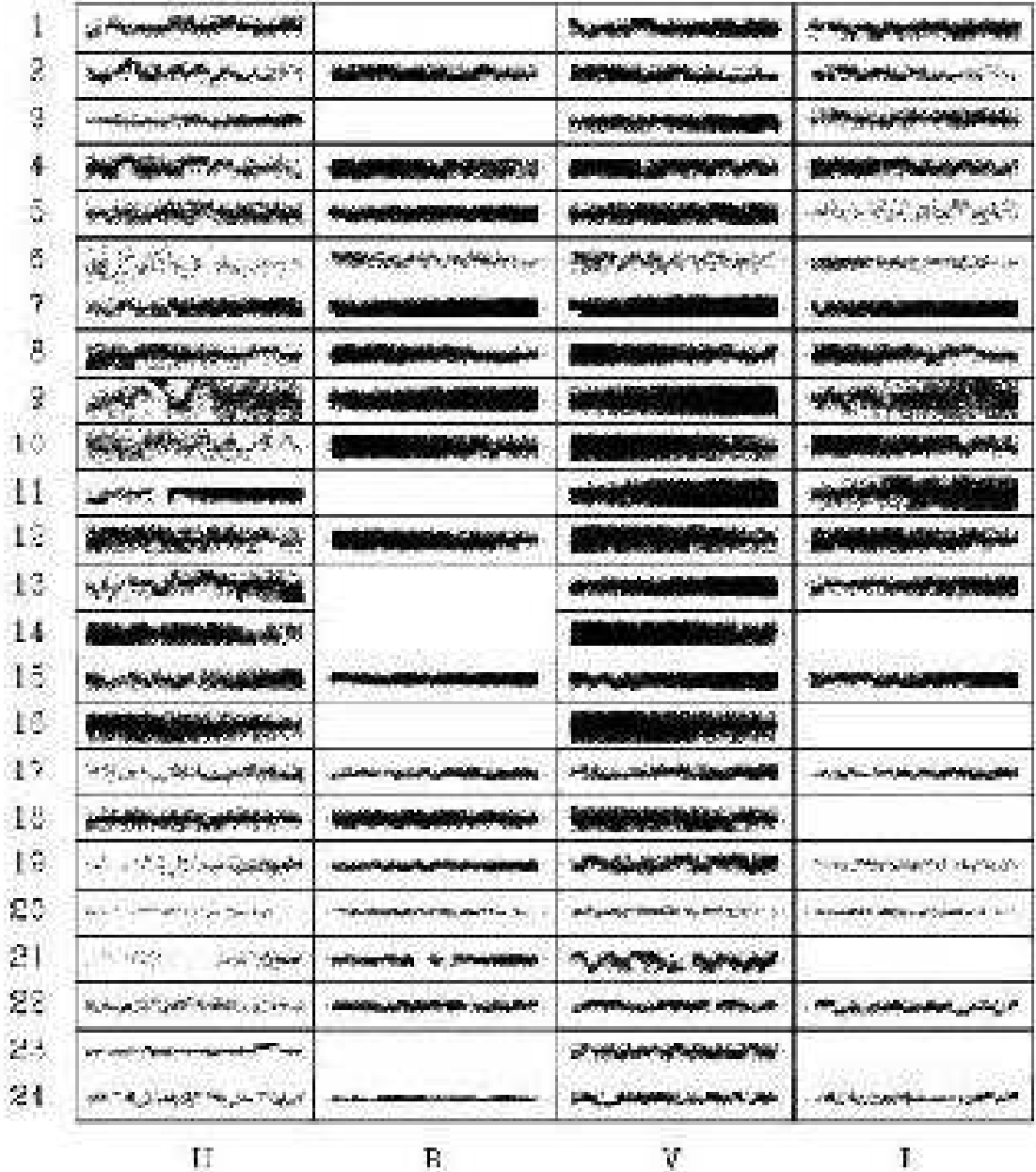}
\figcaption[scancomp.ps]{Photometry comparison of one half of a scan
vs. the other half. Each point represents one star in common. The
numbers to the left indicate the scan numbers (scan 1 is at the
southwest, scan 2 is at the southeast and the scan numbers increase
northward). The filters are labeled at the bottom. In all panels the
vertical axis is $-0.2$ to 0.2 mag and the horizontal axis includes
the entire scan. Empty panels indicate scan/filter combination for
which there was no overlap due to slight misalignment of the
scans. Only stars within 3$\sigma$ of the mean are included in the
comparison, which in highly populated panels leads to sharp horizontal
boundaries for the stellar locus.}
\label{scancomp}
\end{figure}

Our first internal photometric test is provided by the overlap
stars between one half of a scan and the other. We combine all of the subscans
along each side of the scan and then compare the photometry in the overlap region.
Again, because we are comparing the photometry for the
very same stars within the same images, we are 
directly testing PSF model variations. Although
apparently direct, this test is conservative in the sense that we 
are testing the PSF models for stars at the very edges of the frames,
where the PSF models are most poorly constrained. The differences in stellar
photometry across all of the scans where such overlap photometry is available
are shown in Figure~\ref{scancomp}. In a few cases, the overlap region is
unavailable because a scan in a particular passband is offset sufficiently
from the $V$ band scan that the overlap region, defined in the $V$ frame, 
does not contain duplicate
photometry in the other filter. 

Excursions in the photometry
residuals shown in Figure~\ref{scancomp} come in two types: 1) a smoothly
varying fluctuation (see scan 19 in $V$) and 2) a discontinuous jump
(see middle of scan 4 in $U$). The first type of variation is due to
PSF modeling variations. The PSF model is tracking real PSF variations,
which are smoothly varying, differently in the two subscans, resulting in
smooth differences in the photometry. The second type of variation is
due to differences in the aperture corrections applied to the two 
different subscans. Initially, we attempted to correct for clear offsets
such as that seen for scan 4 in $U$. However, we found that these 
applied offsets tended to make the photometry in the entire subscan 
worse. We conclude that there are PSF variations in the subscan that
are not well-modeled near the edge, but that the aperture correction,
which is calculated from stars throughout the subscan, is correct on average,
but incorrect at the edge of the scan in some cases. This subtlety 
illustrates how Figure~\ref{scancomp} provides an exaggerated view of photometric difficulties due to our focus on the edges of the scans. Even so,
the photometric differences 
are typically small (the median of the rms deviations between
different scan halves are 0.032, 0.015,
0.030, 0.028 mag for $U, B, V$ and $I$, respectively). The rms deviation is
$<$ 0.05 mag in both $B$ and $V$ for all scans, and 
is $>$ 0.05 mag for only four $U$ and $I$ scans.

\begin{figure}
\figurenum{5}
\plotone{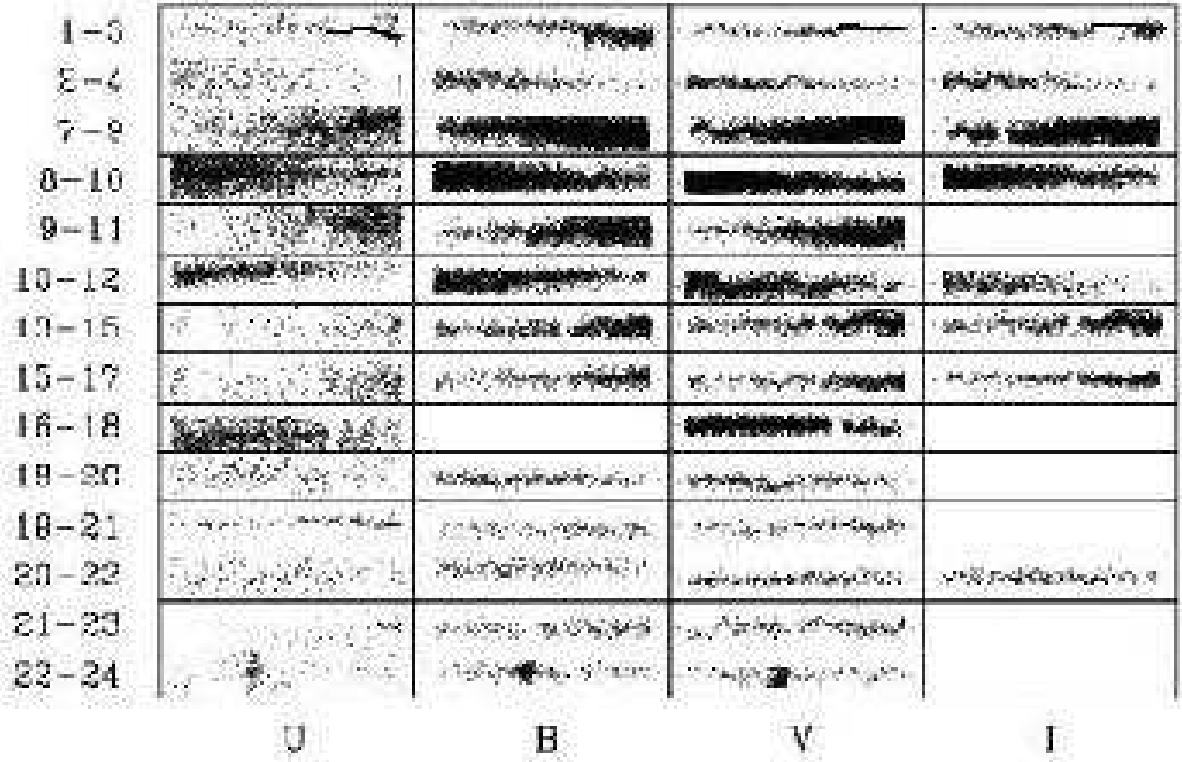}
\figcaption[scans.ps]{Photometry comparison of overlapping scans.
Each dot represents one star in common. The numbers to the left
indicate the scan pair, the filters are labeled at the bottom. In
all panels the vertical axis is $-0.2$ to 0.2 mag and the horizontal
axis includes the entire overlap region.}
\label{scans}
\end{figure}

Our second internal test consists of examining the overlap regions
between different scans. In this case, the images are taken under
entirely different conditions, often in different years with different
detectors. This test provides as good a test of the overall 
photometric accuracy as can be obtained
internally. The magnitude differences along scans are shown for
the four filters in Figure~\ref{scans}. Scan pairs come in E-W pairs
and have the greatest overlap with scans in the N-S directions. Therefore,
scans 1 and 3 overlap along a length of a scan, and scans 2 and 4 overlap.
Some scan pairs that should overlap, such as 5 and 7, do not overlap
because of a slight telescope offset in at least one filter (the gap
between some scan pairs is visible in Figure~\ref{smcimage}.
Again, these plots tend to
exaggerate problems because we are examining the edges of scans, where
the PSF's are most likely to be distorted due to camera misalignments
and tracking errors (as well as the poorly constrained PSF models).
The photometric differences are evidently larger than in the
comparison of scan halves because this comparison is based on different
images of the same stars. The scatter is greater than before, but there
are also some continuous variations (see the $U$ band comparison of scans 10 and
12). Even so, the median rms differences among the scans (comparing 
all overlap stars with $V < 20$) are 0.13, 0.07, 0.06, and 0.05 mag
for $U$, $B$, $V$, and $I$, respectively, These values are comparable 
or smaller than the quoted uncertainties for $V \sim 20$ stars, which dominate the
comparison. We conclude
that to the limit of the data, we have achieved a stable and robust 
photometric system across the survey. 

\begin{figure}
\figurenum{6}
\plotone{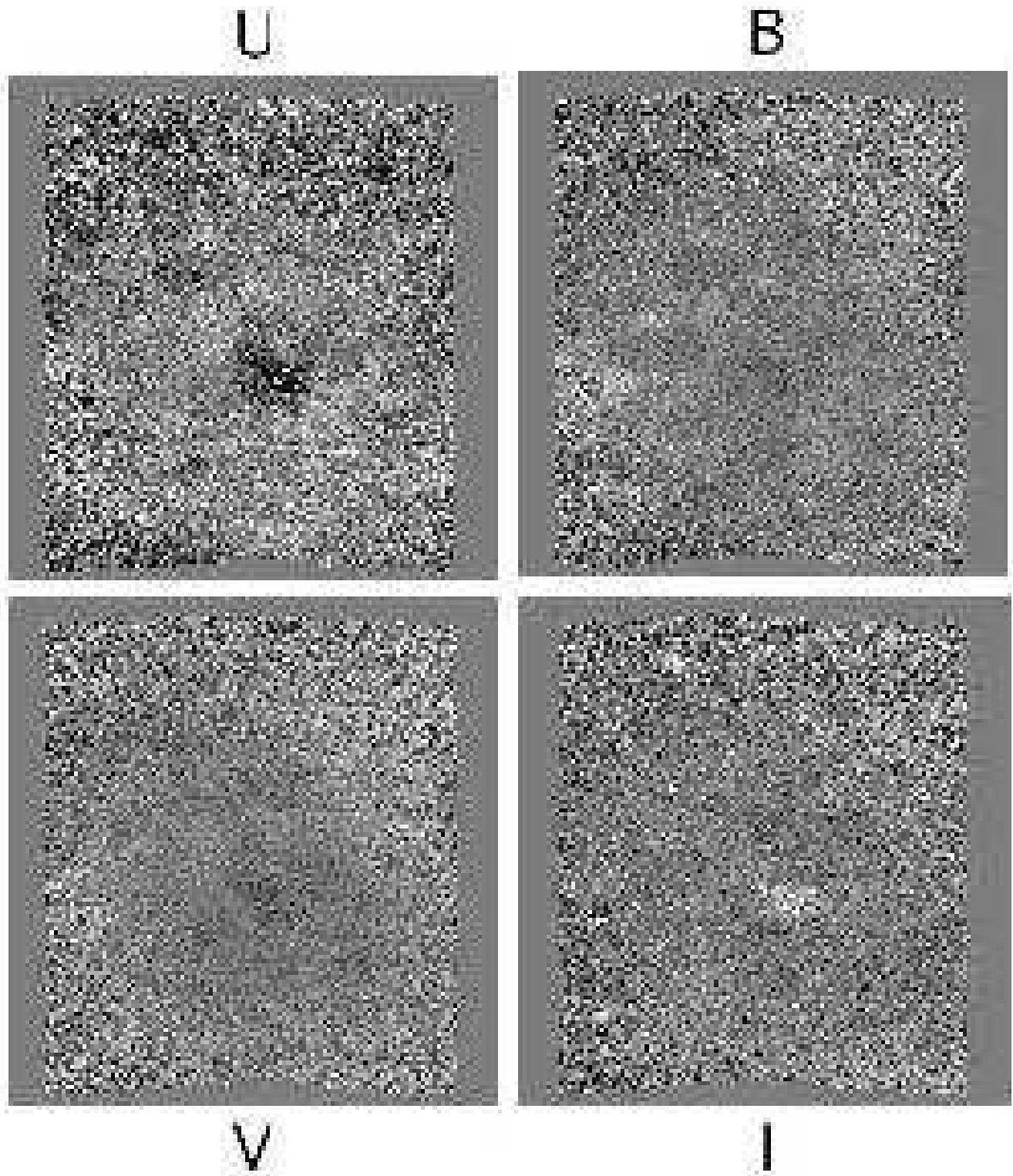}
\figcaption[clumpmos1.ps]{The local median of the red clump magnitude
relative to the global value vs. position across the survey area.
The four frames represent the results from each of the four filters, as
labeled. The greyscales range from $-0.3$ to 0.3 magnitude and all images
have a border beyond the survey area
that is set to the global median, zero.
}
\label{clumpmos}
\end{figure}

Our third internal test consists of using the red clump magnitude to
examine spatial variations in photometry. We calculate the red clump
mean magnitude in 70 $\times$ 70 arcsec boxes using 
stars that have $0.6 < B-V < 1.05$ and
$18.8 < V < 19.8$. Although large-scale variations in the mean magnitude
may truly exists (for example due to a tilt of the SMC relative to a
constant-distance surface; see 
\cite{vdm01} for a demonstration of an analogous effects 
in the LMC), any localized variation, in particular one that traces
scan or subscan boundaries, reveals a problem region.
In Figure~\ref{clumpmos} we show the maps of the residual in the mean
red clump magnitude relative to the global average, for all
filters. There are various important features in these panels. First,
all four panels show increased noise toward the edges because there
are fewer clump stars at large radii from the SMC and contamination
for foreground Galactic stars is proportionally greater. Second, the
$U$ and $I$ band frames in particular show a suspicious feature just to
the southwest of center. This is the most crowded region in the
survey and we suspect that crowding has caused problems for the $U$
and $I$ photometry, which is not as deep as $B$ and $V$.
The $B$ and $V$ photometry does not appear to have serious problems in
this region at the magnitude of the red clump stars. The variations
seen in the mid-left region of $B$ and in the center of the $V$ frame
correspond to less than $\pm 0.05$ mags.  Because these variations are
more irregular (i.e. they do not follow the vertical or horizontal
scan and subscan boundaries) we cannot determine whether these are
real (for example, due to extinction variations) or due to photometric
errors.

In conclusion, our internal tests suggest that there are photometric
uncertainties of the order of a few hundreths of a magnitude between
subscans and scans due to a combination of PSF modeling errors and
calibration uncertainties. With the exception of potential crowding
errors in $U$ and $I$ in the densest region of the SMC, 
the catalog appears to be limited by these
uncertainties and Poisson noise down  to at least $V \sim 20$.

\subsection{External Tests}

We are fortunate to have an array of existing data over substantial
portions of the SMC with which we can further test the photometry
in all four filter bands.

\subsubsection{Comparison to Massey's Catalog}

\cite{massey01} has produced a catalog of SMC photometry ($U, B, V$
and $R$) for bright stars along the central ridge of the SMC and toward
the SMC wing. In particular, because of his interest in the 
upper main sequence, he pays particular attention to obtaining
accurate $U$-band photometry for young SMC stars, 
which is often difficult because of the
lack of blue ($U-B < 0.0$) standards. We discuss various comparisons
between that catalog and ours.

\medskip
\noindent
{\sl Astrometric Accuracy}

\begin{figure}
\plotone{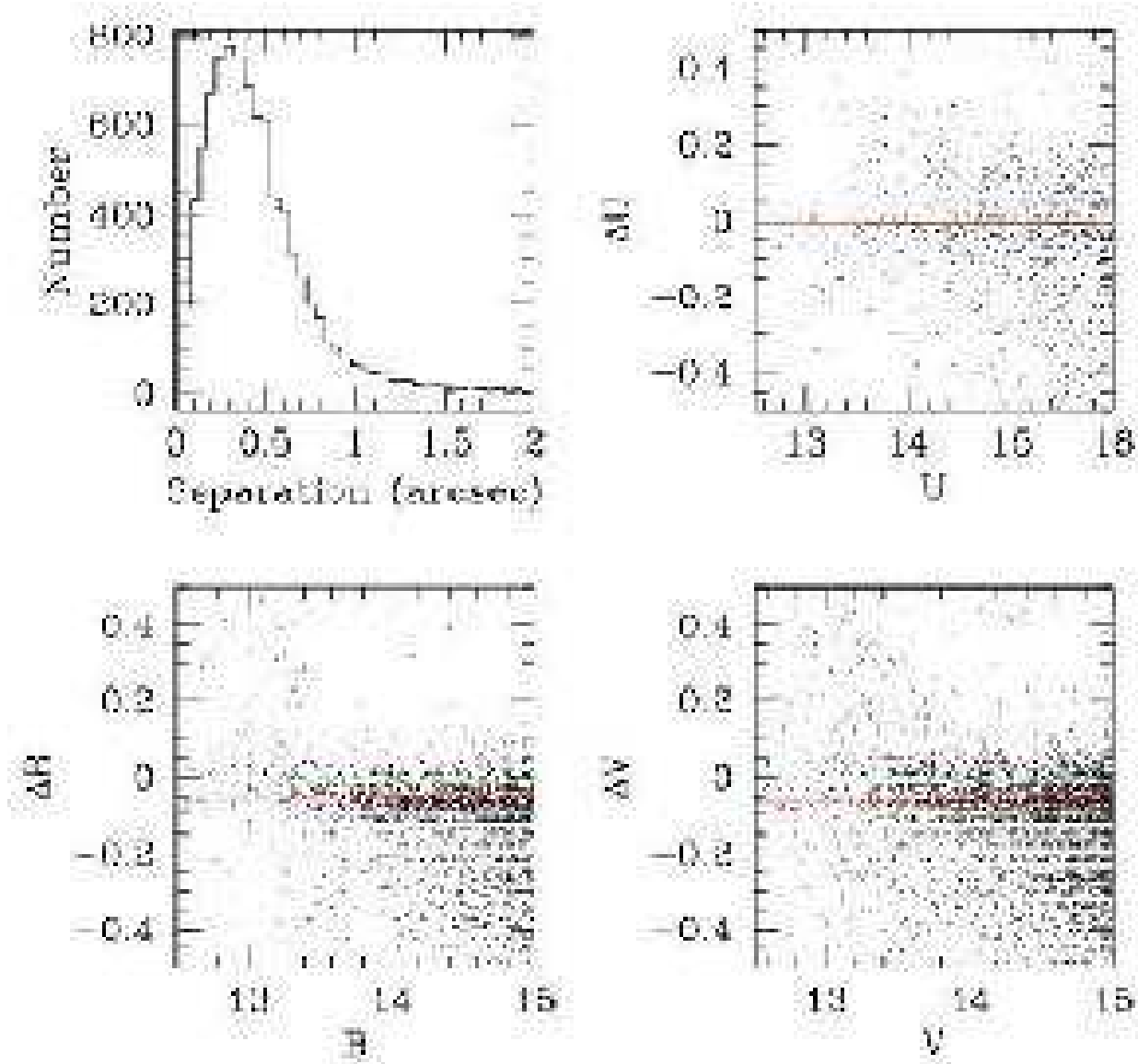}
\figcaption[masseycomp.ps]{The comparison of the UBV photometry from
our survey with that from Massey's (2001) bright star survey.
The upper left panel shows the distribution of matching separations
between stars in the two surveys. The other three panels show the 
comparison of magnitudes with a horizontal line illustrating the 
mean difference over the range of magnitudes for which neither
saturation nor confusion are a significant problem in either survey.}
\label{masseycomp}
\end{figure}

We match stars brighter than $V = 15$ in Massey's catalog to stars in
the MCPS by finding the star in the MCPS within 7.5 arcsec 
of each Massey star that has the closest $V$ magnitude.
The large search radius and the addition of the magnitude criteria has the potential to 
bias us away from picking the nearest star in projection. However, the
distribution of separations (Figure~\ref{masseycomp}, upper left) 
shows a strong peak
toward matches with an offset that is $< 1$ arcsec (the distribution peaks at
$\sim 0.3$ arcsec). Although all matched stars are included in 
Figure~\ref{masseycomp}, the offsets for stars with $V < 13.5$ are somewhat
larger and the reason for this is discussed in below.
The astrometric correspondence between these two catalogs is
excellent -- certainly sufficient for spectroscopic follow-up
of these stars.

\medskip
\noindent
{\sl Photometric Zeropoint Comparison}

In Figure~\ref{masseycomp} 
we plot $\Delta m$ vs. $m$ for matched stars in $U, B,$ and $V$, where
$\Delta m$ is defined to be Massey's magnitude minus the MCPS
magnitude. The bulk
of the stars lie in a broad horizontal locus in these plots, with a flaring
at $V < 13.5$ toward positive residuals, and a flaring at 
$V> 14.5$, toward negative residuals. 
Limiting the comparison to  stars with positional offsets $\le 0.7$ arcsec  
and $0.35 < B-V$ for the $U$-band panel
(see below for discussion
of the color cut) and $13.5 < V < 14.5$ for $B$ and $V$ (eliminating
the brighter magnitudes, where the MCPS has problems, and fainter
magnitudes, where Massey's data appear to have problems; see
below) we calculate the mean offsets relative to $\Delta m$ = 0 to be
$-0.057$ (after correction, see below), $-0.063,$ and $-0.057$ mag for 
$U, B,$ and $V$
respectively. These offsets are consistent with the photometric
zeropoint errors of the studies, although slightly larger than
expected. Fortunately, the colors, which are more sensitive to
photometric zeropoint errors, are consistent between the two
studies. We suspect that the systematic trend to negative mean
residuals is at least in part due to the flaring toward negative
residuals for individual stars at fainter magnitude, 
rather than a true photometric zero point difference.
This flaring, as we show below, is due to
crowding. An important consideration in this comparison, especially
at the faint end, is that Massey's magnitudes are aperture magnitudes
within a large (8.1 arcsec) aperture. 

\begin{figure}
\figurenum{8}
\plotone{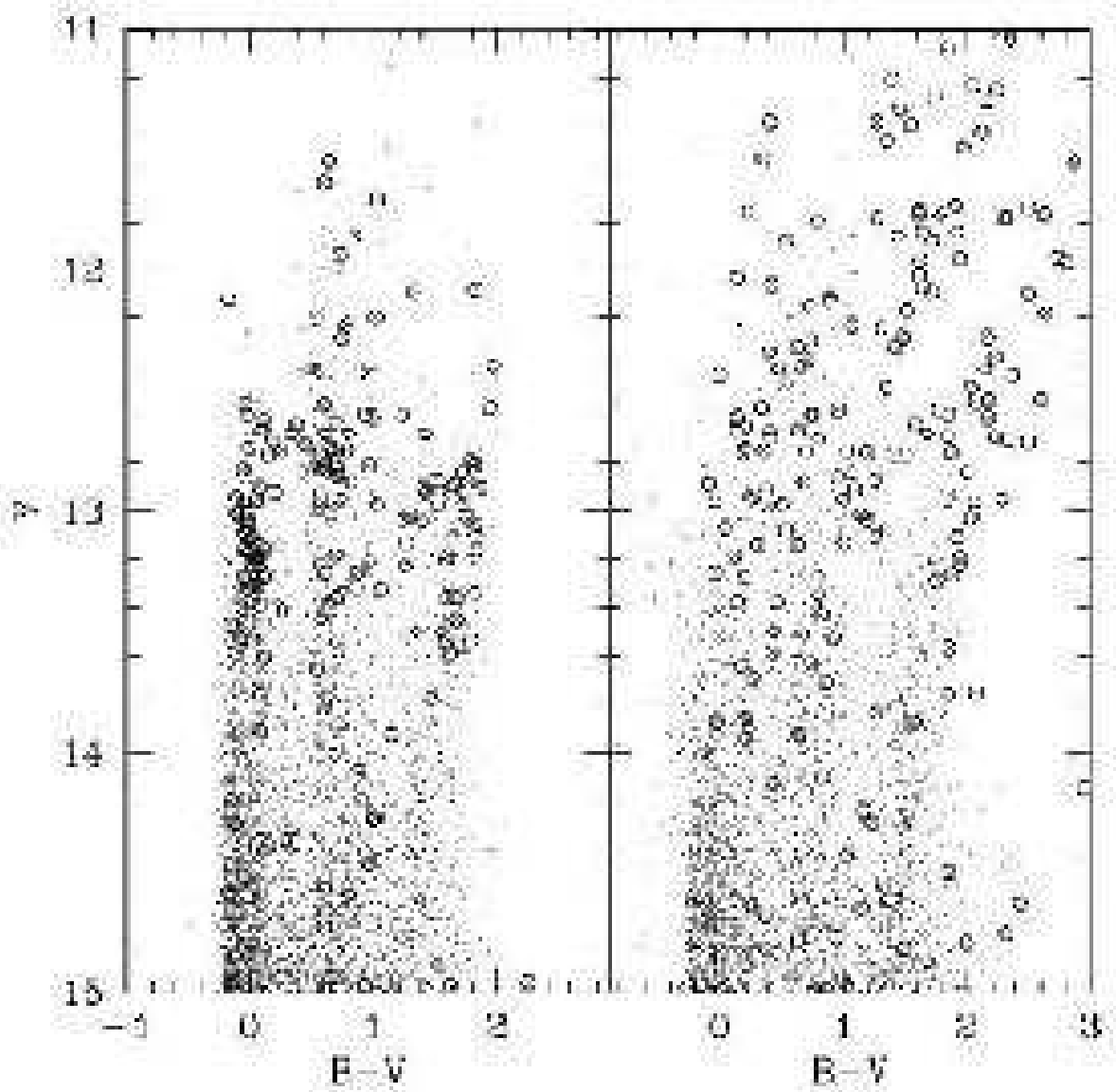}
\figcaption[masseybright.ps]{The bright star color-magnitude
diagram using the data from Massey's catalog on the left and from
the MCPS on the right, for stars that have been matched between
the two surveys. The open circles represent stars with $\Delta V > 0.2$ 
mag and positional discrepancies 
$<$ 0.7 arcsec. }
\label{masseybright}
\end{figure}

As already mentioned, in addition to the zeropoint offset of the horizontal 
locus, other features are visible in these diagrams. First, there is a diagonal
flaring toward the upper left in the $B$ and $V$
panels of Figure~\ref{masseycomp}. 
These are stars from the \cite{massey01} catalog that are matched to
a ``brighter" star in our catalog.  Because this problem appears only at
bright magnitudes, we suspect it is due to poor PSF fitting of bright
stars in the MCPS.  In Figure~\ref{masseybright} we show the
color-magnitude diagrams for $B-V, V$ and highlight the stars with
$\Delta V > 0.2$ for both Massey's catalog and the MCPS. It
is evident from the tightness of the upper main sequence and red
supergiant envelope for these stars that Massey's photometry
is superior at these bright magnitudes (the sequence visible between
the main sequence and supergiants in the left panel are the foreground
dwarfs).  We conclude that stars in
our catalog that are brighter than 13.5 in $B$ or $V$ are prone to
substantial photometric uncertainty.  Therefore, we have replaced the
photometry and astrometry in our catalog for stars brighter than this
limit with that of \cite{massey01}. In replacing our photometry
with Massey's, we apply the mean photometric offsets found between the
MCPS and Massey's catalog to Massey's
data. Because Massey's catalog
does not cover the entire area of our survey, not every star in the
MCPS with $B, V < 13.5$ has corrected photometry. To indicate which
stars do have corrected photometry the catalog includes a flag (see
\S5 for a discussion of all of the quality flags). 809 stars are
affected by this correction.

\begin{figure}
\figurenum{9}
\plotone{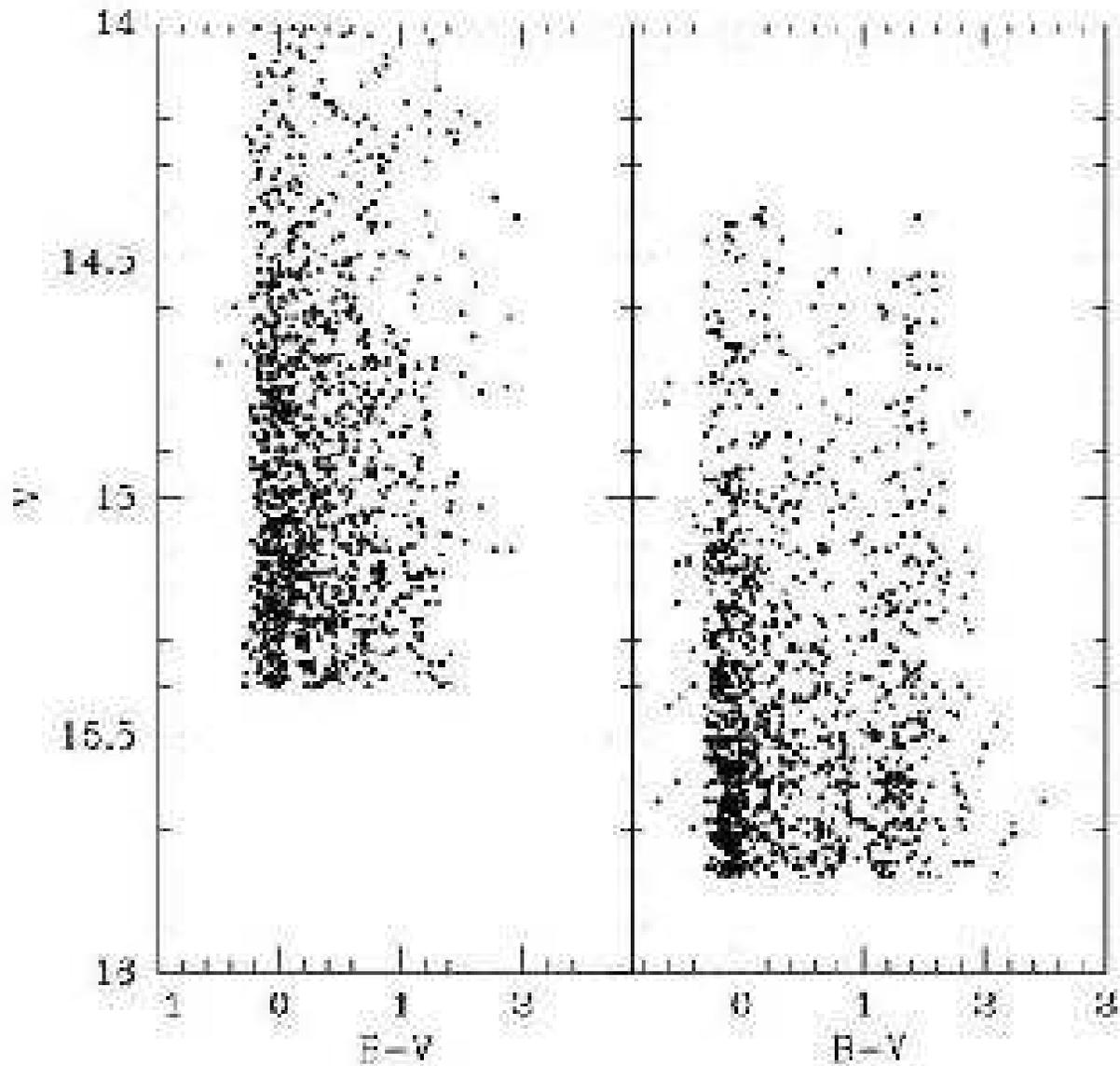}
\figcaption[masseyfaint.ps]{The color-magnitude diagram of
fainter stars in common between the 
Massey survey on the left and the MCPS
on the right that have $\Delta V < -0.3$
mag (the faint tail). The vertical (magnitude) offset between the stars
in the two panels arises because we only include stars with $\Delta V
< -0.3$.}
\label{masseyfaint}
\end{figure}

The second systematic feature in these diagrams is the asymmetric
distribution of $\Delta m$ toward fainter magnitudes.
In this case, stars from the \cite{massey01} catalog have been matched
to  fainter stars in the MCPS. Such a mismatch could occur
if Massey's catalog is beginning to be susceptible to 
unresolved blending, which is likely given the 8.1 arcsec
aperture used for the aperture photometry.  By comparing color-magnitude
diagrams of stars with residuals $\le -0.3$ (Figure~\ref{masseyfaint}),
we find that in contrast to the situation with positive $\Delta m$,
our data produce sharper main sequence and red envelope sequences.
We conclude that at the faint magnitude limit
of the comparison range our data are superior.

\medskip
\noindent
{\sl Photometric Color Term Comparison}

In addition to zeropoint discrepancies, we use the Massey's
catalog to check for incorrect or missing color terms in our
photometric solution. This issue is particularly important for
the $U$ band because there are few blue calibration stars 
available in the standard fields. \cite{massey01} has taken particular
care to include a significant number of blue upper main sequence
stars in the calibration, so we have reason to believe that
his calibration in this regime is superior to ours. 

\begin{figure}
\figurenum{10}
\plotone{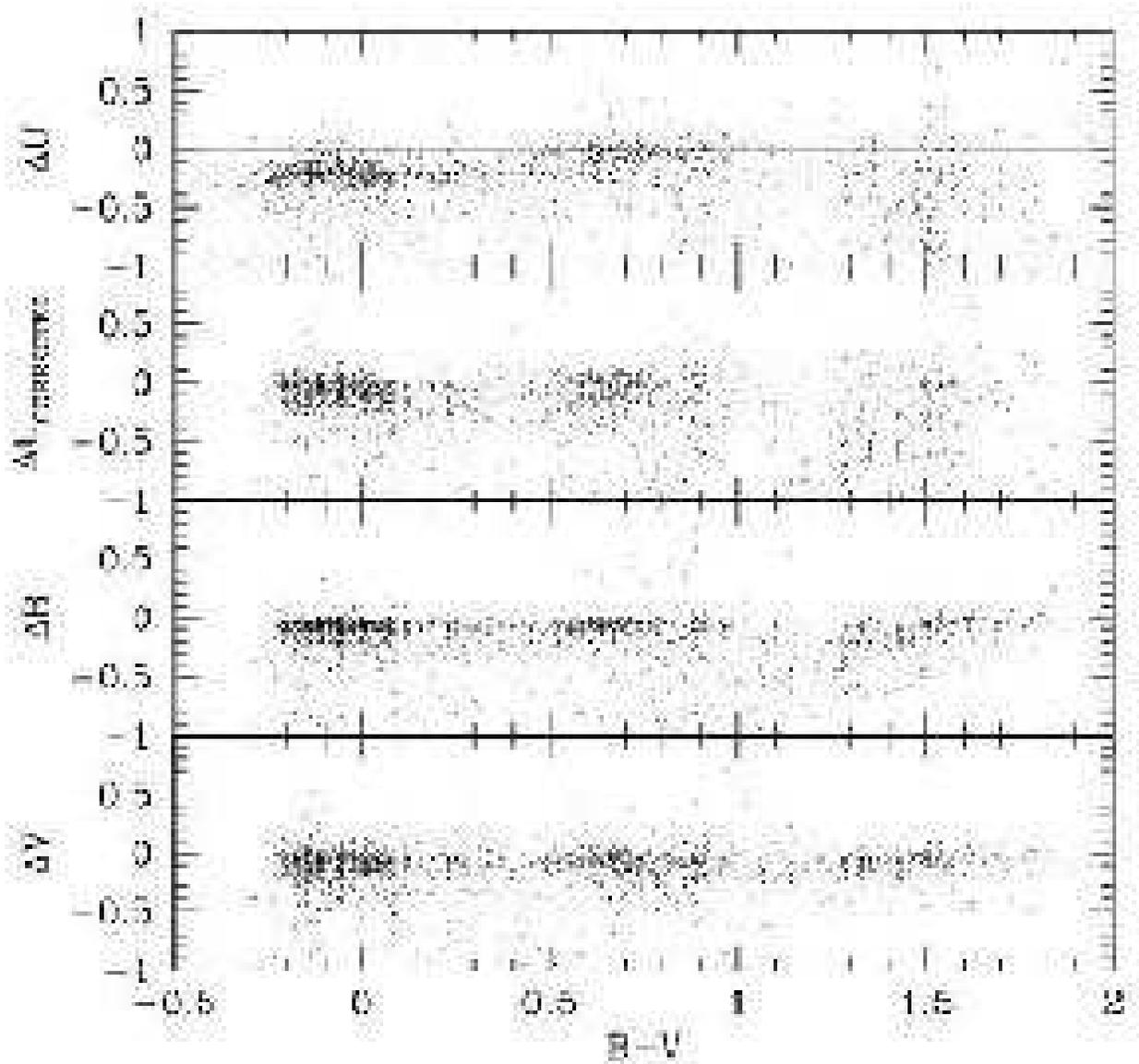}
\figcaption[masseycolor.ps]{The comparison of magnitudes 
as a function of stellar color 
between the Massey survey and the MCPS. The upper panel shows the results
from the original $U$ band photometry, the second panel shows the
corrected $U$ band photometry. }
\label{masseycolor}
\end{figure}

The comparison of magnitudes for matched stars as a function of their
$B-V$ color is shown in Figure~\ref{masseycolor}. There are no grossly
discrepant color terms in the $B$ and $V$ photometry, but the $U$ band
photometry shows a noticeable shift ($\sim 0.2$ mag) 
the bluer and redder stars. 
Because we do not know whether the discrepancy lies in the bluer or redder
stars, we apply the simplest (linear) correction to the $U$ band magnitudes 
that resolves this problem and minimizes the color
differences between the MCPS and Massey's catalog.
This correction involves increasing the magnitudes
by 0.05 mag for stars with $0.4 < B-V < 1.0$ and decreasing them
by 0.05 mag for stars with $0.4 > B-V$.
We show the corrected photometry in Figure~\ref{masseycolor}. 
The only other potential missing color term is suggested by a
feature in the $B$-band photometry at $B-V \sim 1.1$ to 1.3. This is a
sufficiently sharp feature (the $B$-band photometry returns to $\Delta
m = 0$ for $B-V > 1.4$) that we do not correct for it and suspect that
it is the result of beating between a slight difference in the filter
transmission curve of the two studies and a spectral absorption
feature in stars of this surface temperature. At worst, it may result
in a $\sim$ 0.1 mag mean error in the MCPS $B$ photometry of stars of
this color.

\medskip
\noindent
{\sl Verifying the Uncertainties}

A critical component of catalogs such as the MCPS is the uncertainty
estimates. For example, the error distribution plays a key role in
synthesizing color-magnitude diagrams to recover the star formation
history. Because the uncertainties come from a wide range of phenomena
(calibration, PSF modeling, data quality variations, crowding), it is
difficult to know whether the quoted uncertainties accurately reflect 
all of these. In particular, we need to verify that we understand the
uncertainties that cannot be recovered with artificial star tests, such
as those in the photometric calibration.

\begin{figure}
\figurenum{11}
\plotone{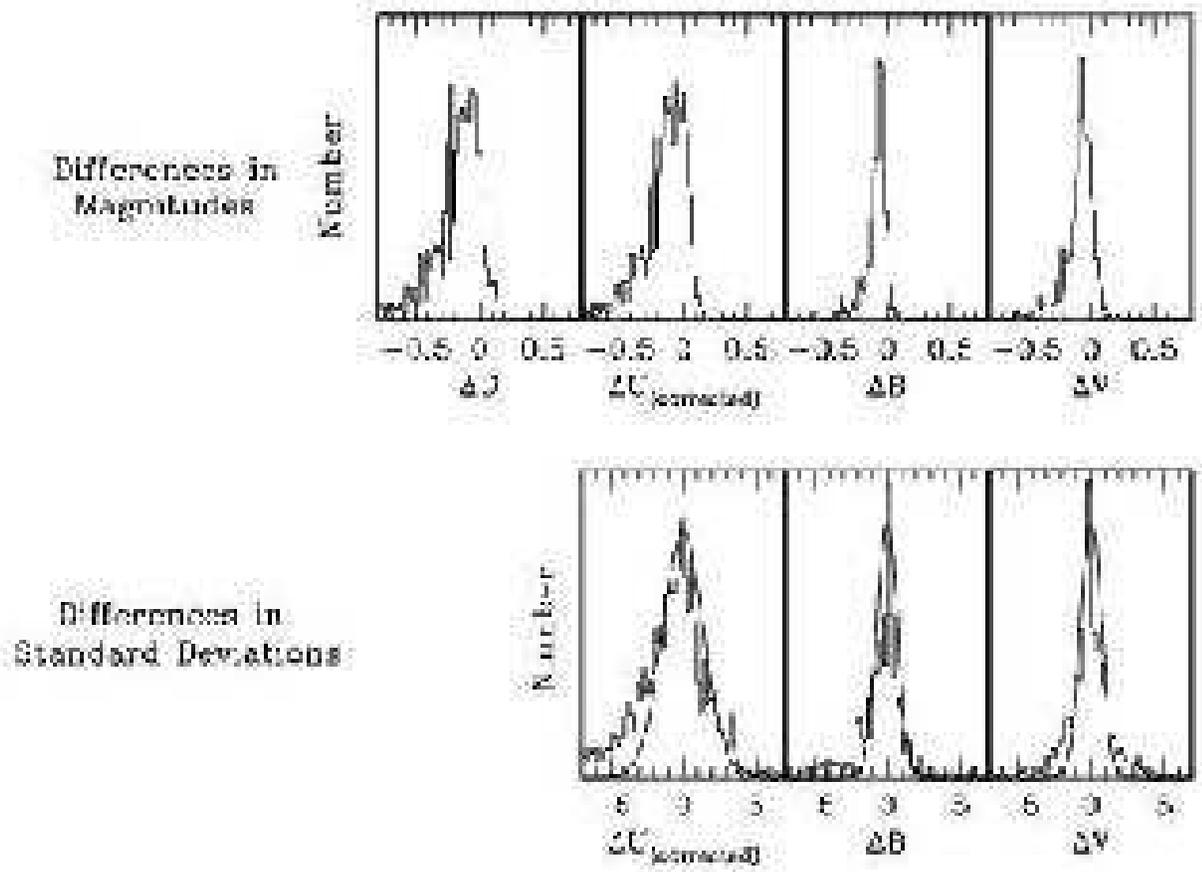}
\figcaption[masseyhist.ps]{Histogram of magnitude differences between 
the Massey survey and the MCPS.
Upper panels show the distributions of 
magnitude differences, lower panels show the distribution of
magnitude differences 
in units of standard deviations (using the uncertainties associated with
each star in the two catalogs). The curves plotted in the lower
panels are Gaussians of $\sigma = 2, 1, 1$ for $U, B,$ and $V$
respectively.}
\label{masseyhist}
\end{figure}

In Figure~\ref{masseyhist} we plot the distribution of magnitude
differences for stars in Massey's catalog and the MCPS that
are in a magnitude range that is minimally affected by saturation
and crowding ($B$ and $V > 13.5$, $U < 15$, and $B$ and $V < 14.2$)
and that have matches 
with separations less than 1 pixel (0.7 arcsec). The upper set of
panels shows the $\Delta m$ distribution in magnitudes, the lower set shows 
the distribution in terms of standard deviations, propagating the uncertainty
estimates from both catalogs.
The histograms of $\Delta m$ for both $B$ and $V$ have a core that is actually
slightly narrower than the Gaussian prediction, and 
wings that are larger, particularly in $V$. We suspect that the wings are due to zeropoint 
fluctuations between subscans that are somewhat larger than that suggested
by the statistical errors of the photometric calibration 
(we will see more evidence of this effect later). 
The $U$ band data is fitted by a Gaussian that is
twice as wide as expected, suggesting that the uncertainties are
significantly underestimated. Again, we suggest that in large part
this reflects zeropoint fluctuations among scans. For example, 
additional random fluctuations
of 0.03 mag in the photometric zeropoints from subscan to subscan
are sufficient to account for the widened distribution. 
This fluctuation
in the zeropoints is consistent with the results of the 
internal comparison discussed previously.

\subsubsection{Comparison to the OGLE Catalog}

\medskip
\noindent
{\sl Zeropoint Calibration}

The most extensive photometric data set available for comparison to
our MCPS is that provided by the OGLE group (\cite{udalski98}).  They
have measured $B$, $V$, and $I$ magnitudes for stars along the central
ridge line of the SMC. Again, we match stars between our catalog and
theirs, and identify a magnitude regime where there appear to be no
systematic problems. For this comparison we restrain the comparison to
$B$ and $V >13.5$, a faint limit of 15 for $B$ and $V$, and 17 for
$I$, and $|\Delta m| < 0.2$. The mean photometric offsets are 0.011,
0.038, and 0.002 mag in $B$, $V$, and $I$ respectively. These are all
within the internal uncertainties of our zeropoint calibration and in
the opposite sense as the zeropoint differences resulting from the
comparison to the \cite{massey01} catalog. Again, we suggest that at
least some part of the mean offsets are not due to photometric zero
point errors, but to the systematic trend at fainter magnitudes (which
is not due to calibration errors but rather to crowding).
We conclude that our average, global photometric zeropoints are
good to better than 0.03 mag.

\begin{figure}
\figurenum{12}
\plotone{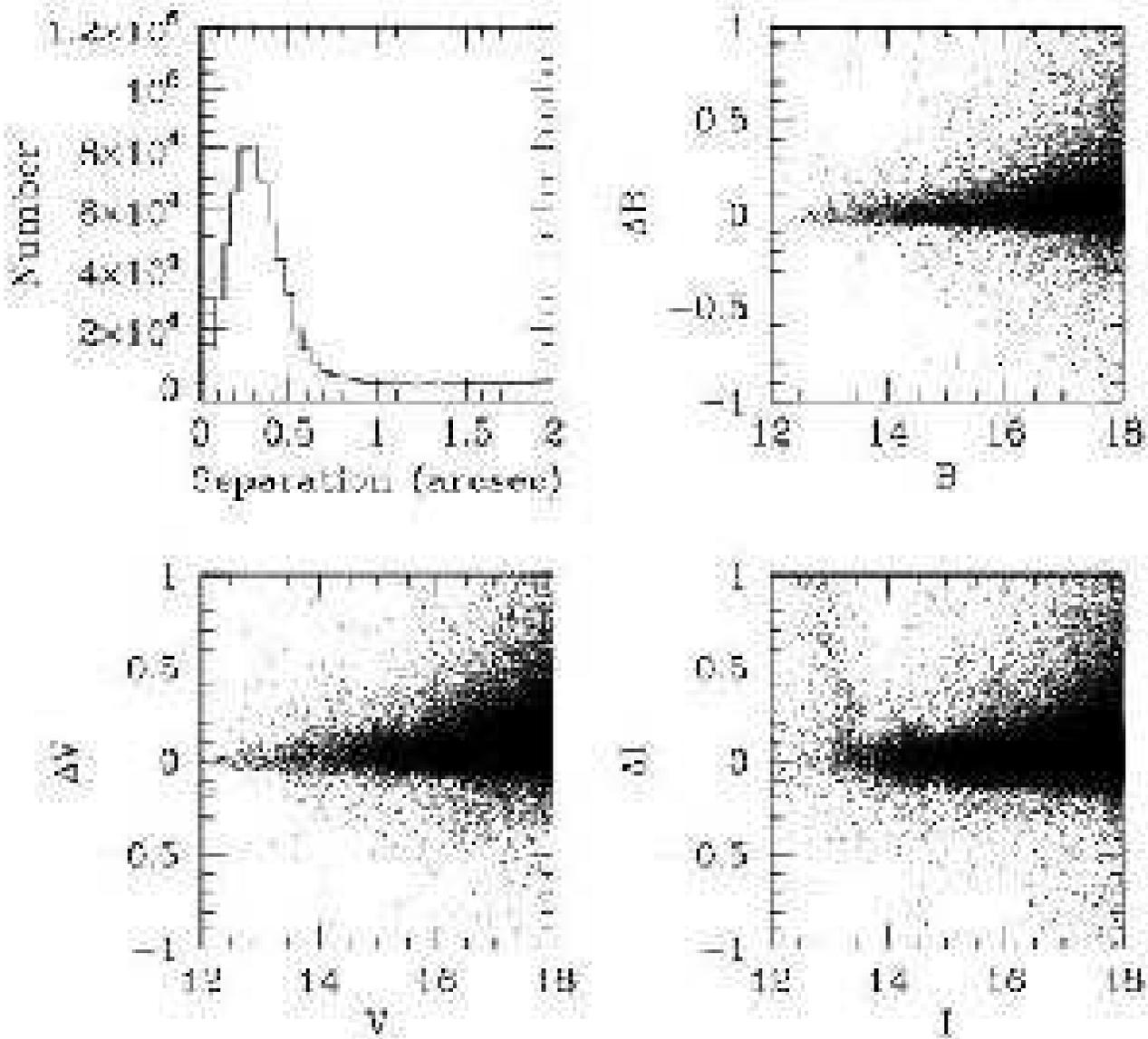}
\figcaption[oglecomp.ps]{The comparison of the $BVI$ photometry from
our survey with that from the OGLE survey.
The upper left panel shows the distribution of matching separations
between stars in the two surveys. The other three panels show the 
comparison of magnitudes.}
\label{oglecomp}
\end{figure}

\begin{figure}
\figurenum{13}
\plotone{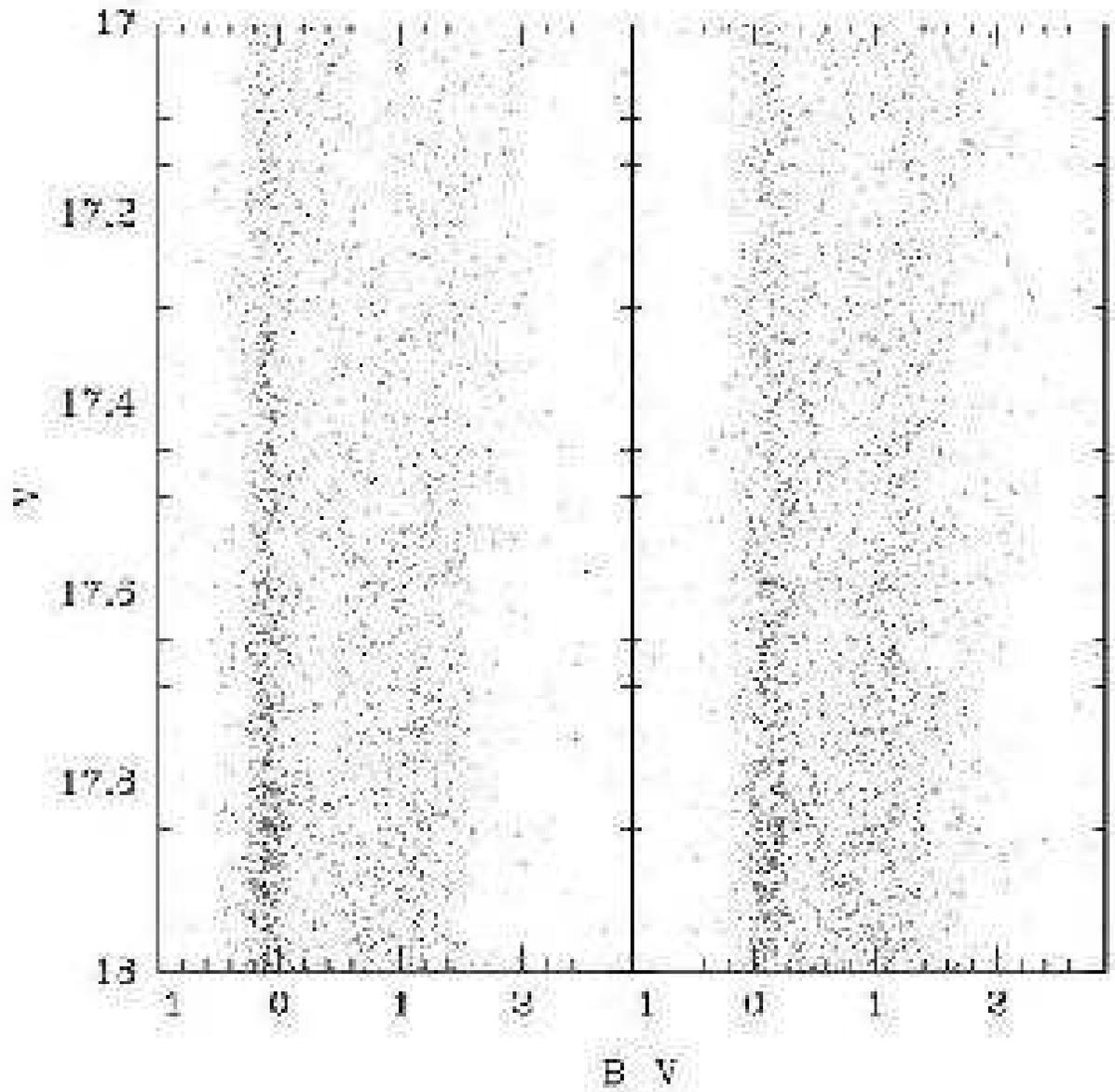}
\figcaption[oglefaint.ps]{The color-magnitude diagram of
fainter stars in common between the OGLE survey (left)
and our survey (right) that have $\Delta V>$ 0.3
magnitudes (the faint tail). }
\label{oglefaint}
\end{figure}

In Figure~\ref{oglecomp} we show the same astrometric and photometric
comparison that we did previously for the \cite{massey01} catalog,
except this time in $B,$ $V$, and $I$. We find again that the
astrometric agreement is good to a fraction of a MCPS pixel. 
Regarding the magnitude comparison
we find patterns similar to those seen previously. In the
$I$ band comparison there is an upward tail at bright magnitudes.
This is due to poorly modeled bright stars in the MCPS. These tails are not visible 
in the $B$ and $V$ comparisons because the photometry of  stars with $B$ and $V < 13.5$
was already replaced with the \cite{massey01} photometry. We now replace
the $I$ band 
photometry of suspect, bright stars ($I < 13.5$) and an associated
flag is set in the catalog --- see \S5). There are 1095 stars that
are affected by this correction. We also see the asymmetry in residuals
at faint magnitudes, however this time the tail is toward positive
residuals. Comparing color-magnitude diagrams for stars with
$\Delta m > 0.3$ we find that the sequences appear tighter using
the OGLE photometry (Figure~\ref{oglefaint}). 
We conclude that the OGLE photometry is superior at the
fainter  magnitude levels. 
Nevertheless, the uncertainties in the MCPS catalog do include an
approximation of the photometric uncertainties due to crowding. The 
$\Delta m$ distribution for the fainter stars is consistent, with
the exception of a low-level tail toward positive residuals, with
a Gaussian of unit dispersion and the additional scatter due to
subscan-to-subscan photometric variations discussed earlier. 
Although
the measurement uncertainties do reflect the crowding problems,
any analysis that is sensitive to the details of the these errors
should use artificial star tests (see \cite{hz01} for a 
discussion of those tests).

\medskip
\noindent
{\sl Color Terms}

\begin{figure}
\figurenum{14}
\plotone{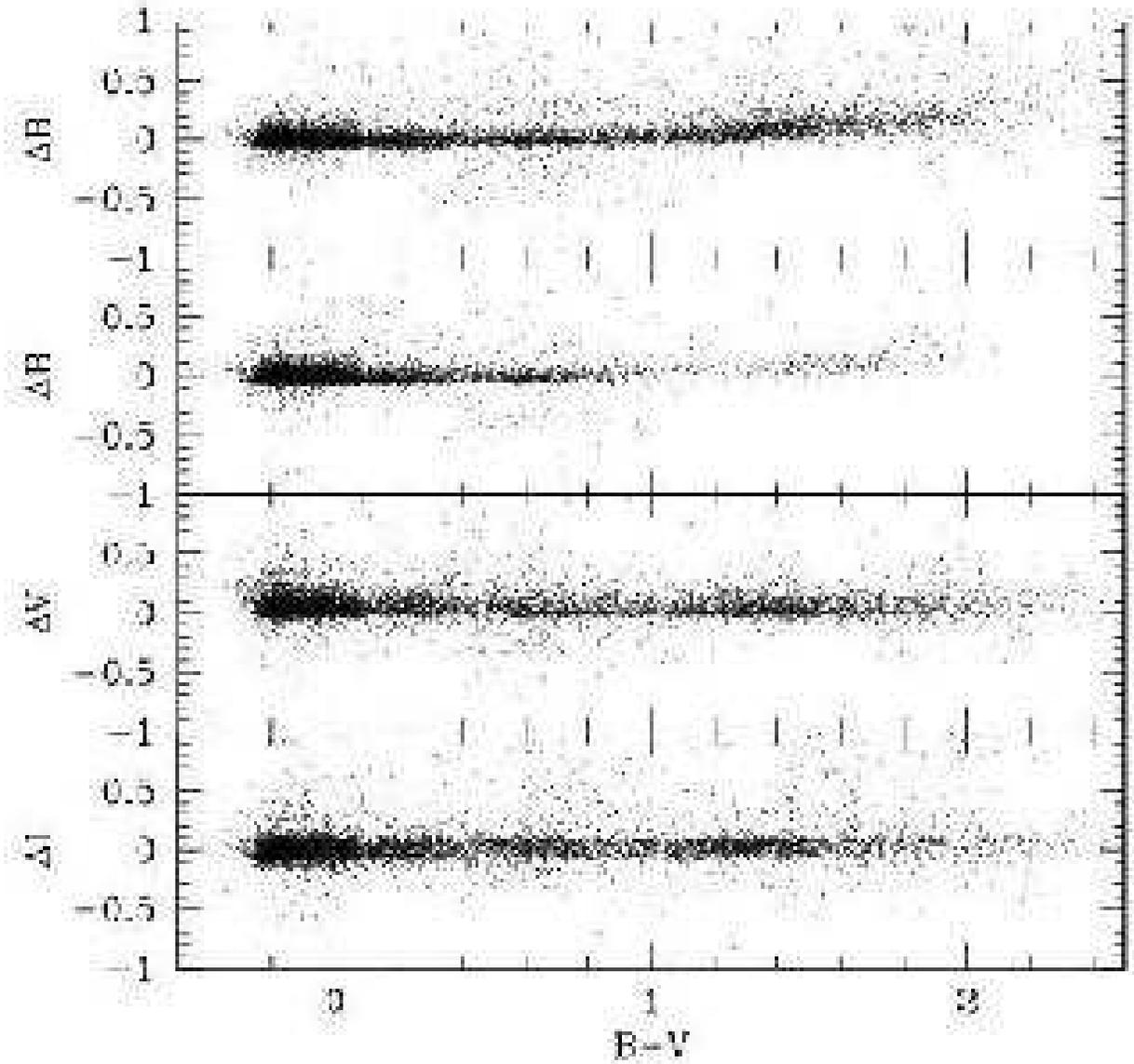}
\label{oglecolor}
\figcaption[oglecolor.ps]{The residual magnitudes of stars matched
between the OGLE and MCPS catalogs as a function of $B-V$ color.
The uppermost panel shows the $B$ photometry for the same sample
used in the lower two panels, while the second panel shows the
comparison only for stars satisfying a brighter $B$ limit. In both
of the upper two panels the color-dependency of the $B$ band 
residuals is evident.}
\end{figure}

Using the matched stars, we examine whether there are color-term variations
between the two studies. From Figure~\ref{oglecolor}, we conclude that the
$V$ and $I$ photometry have no residual color-term dependences. In
contrast, the $B$ photometry appears to have a systematic residual at red colors.
Limiting the comparison to bright $B$ magnitudes, to avoid including
many stars with large photometric residuals due to 
crowding, we still find the systematic color-dependent residual. Such 
a color-term is feasible because there are few standard stars at
these extreme colors (Figure~\ref{stands}). However, the comparison to 
Massey's catalog does not show such a problem (Figure~\ref{masseycolor}). 
In that comparison, there is a photometry difference
at $B-V \sim 1.2$, but by $B-V = 1.5$ the $B$ photometry between the 
two studies agrees well (while
the MCPS and OGLE photometry disagree by 0.1 mag at $B-V = 1.5$). 
Because the comparison of the MCPS with these two surveys is inconclusive,
and we have no further reason to suspect our photometry, 
we do not correct for the color term but 
caution that for the reddest stars there may be a $B$-band systematic
error of $\sim 0.1$ mag in either our data or the OGLE data.

\medskip
\noindent
{\sl Verifying the Uncertainties}

\begin{figure}
\figurenum{15}
\plotone{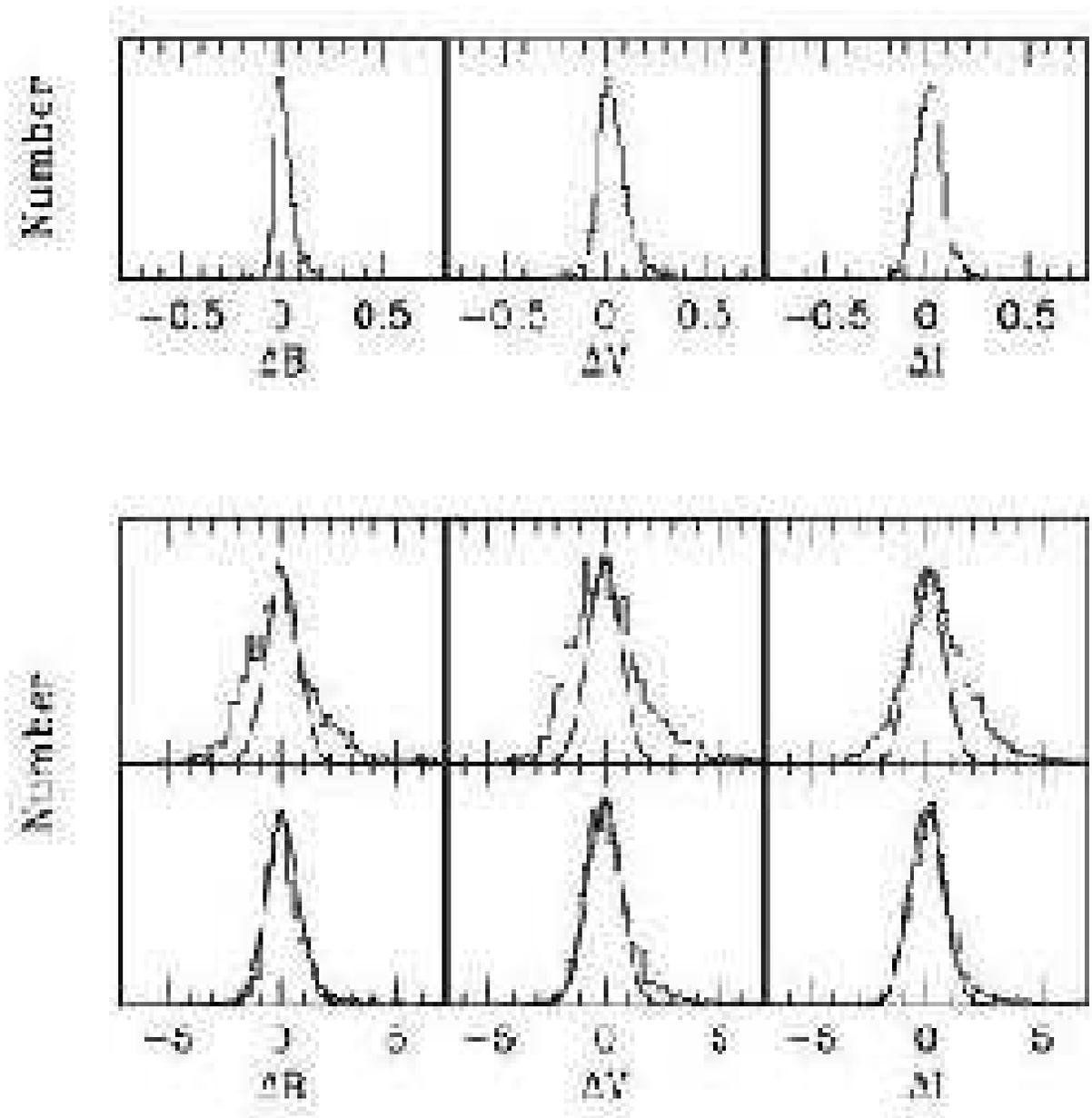}
\figcaption[oglehist.ps]{The distribution of magnitude
differences between the OGLE survey and the MCPS.
In the top panel we show the distribution in terms of magnitude
differences for the three bands in common. In the middle set of
panels is the distribution in terms of standard deviations and
a Gaussian of unit dispersion is overplotted for comparison. In
the lowest set of panels we show the same except we have added
a dispersion of 0.03 mag to all uncertainties to model the 
scatter expected in the zeropoints from scan to scan.}
\label{oglehist}
\end{figure}

Similar to our comparison to Massey's catalog, we plot
the distribution of magnitude differences for matched stars
in units of magnitudes and 
standard deviations (Figure~\ref{oglehist}) for stars with
astrometric differences $<$ 0.7 arcsec, $B$ and $V > 13.5$,
$B$ and $V < 15$, and $I < 16$.
A Gaussian of width corresponding to the
propagated uncertainties with an additional random photometric
zeropoint uncertainty of 0.03 magnitudes agrees well with the distributions
in all three bands. We conclude that the error estimates, with this
small additional term due to uncertainties in the scan photometric 
zeropoints, describes the total uncertainties well at magnitudes where
crowding does not play a role. 

\medskip
\noindent
{\sl Spatially Varying Photometric Comparison}

\begin{figure}
\figurenum{16}
\vskip -9in
\plotone{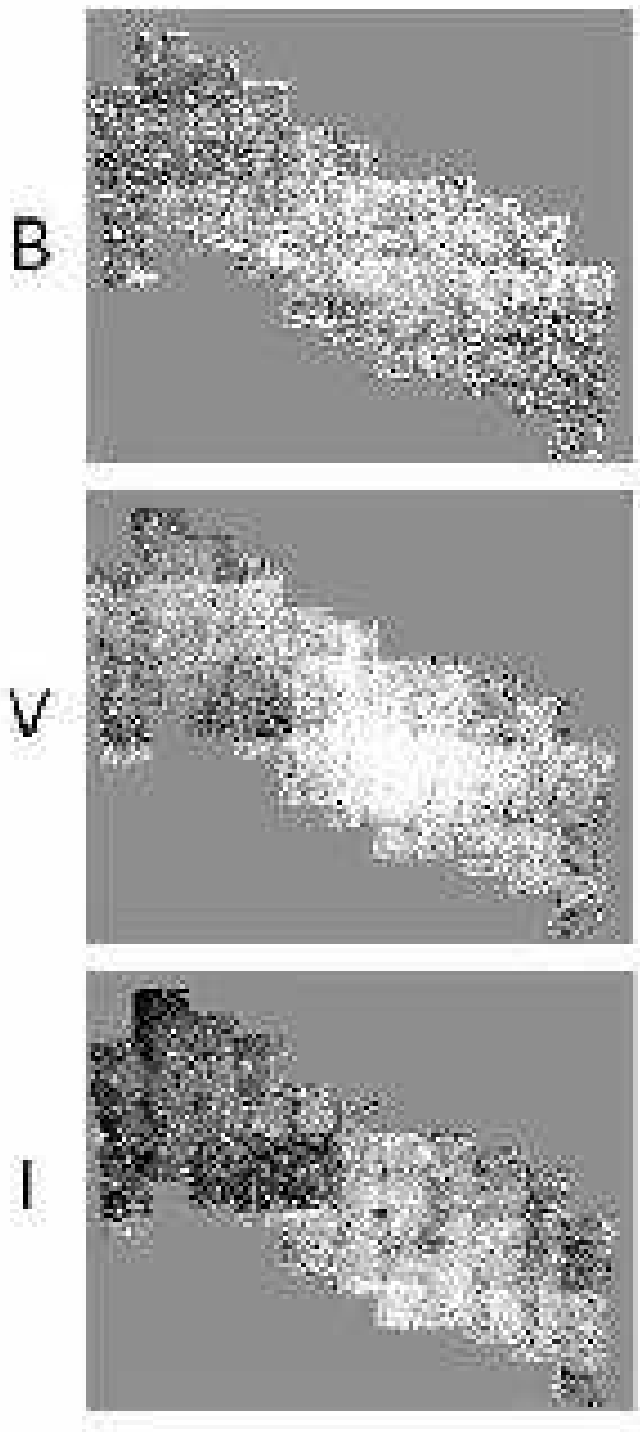}
\figcaption[oglemos.ps]{The spatial distribution of median
magnitude differences
between the OGLE survey and the MCPS for the
three bands in common ($B$, $V$, and $I$). The OGLE survey 
concentrated on the ridge of the SMC and does not include the
entire MCPS survey area. The grayscale covers the range $-0.1 < \Delta m
< 0.1$.}
\label{oglemos}
\end{figure}

The large sample of stars in common between MCPS and OGLE enables
us to compare the photometry spatially over the area in common, and so
determine whether there are problems on the scale
of an individual scan or subscan (either in the MCPS or OGLE).
In Figure~\ref{oglemos} we present maps of the photometric offsets
(median magnitude differences within square ``pixels" for the matched
stars used earlier to study global differences) for $B$, $V$, and
$I$. Two types of spatial patterns are visible. The first consists
either of vertical or horizontal striping. This striping is due to
photometric errors in scans (MCPS has horizontal scans in this
orientation, OGLE has vertical scans).  Because different MCPS scans,
even in the same filters, may come from runs separated by several
years, it becomes difficult to obtain photometry that agrees to better
than the internal calibration errors (typically $\sim 0.03$ mag). A
horizontal edge is visible in the $I$-band, near the middle left of
the image. Vertical edges are visible in the right half of the
$I$-band panel. The difference across these edges is $\pm 0.05$ mag
from the mean for the most noticeable edges.  The second spatial
pattern is the increasing discrepancy (especially in the $V$ band)
toward the SMC center. This is almost certainly the effect of
increasing crowding.  At its worst, this offset appears to be 0.08
mag.

As the crowding increases beyond the ability of the data and software
to disentangle, it becomes increasingly likely that fainter stars
will contaminate the photometry of brighter stars (and that an
increasing fraction of fainter stars will be lost). Therefore,
the detected stars will become increasingly brighter (which is what
is observed when comparing  MCPS data to OGLE data in the crowded
regions). OGLE is superior in this respect  for several reasons:
1) their template images are taken in seeing as good as 0.8 arcsec,
2) their pixel scale (0.417 arcsec pixel$^{-1}$) is much better suited to
the best seeing episodes than ours, and 
3) they use repeat observations to
cull unreliable photometry.
In particular, our $V$ scan of the central region is not optimum. Obviously,
higher quality data is desirable, but for some applications
(such as the generation of synthetic color-magnitude diagrams) this
effect can be included in the simulations and results only in a loss
of information, not in a systematic error. Alternatively, 
investigators interested in the densest regions of the SMC may want
to construct a hybrid catalog that uses OGLE data along the
SMC ridge and MCPS data beyond the ridge. Users of the MCPS
should be aware that there is a bias toward measuring a 
brighter magnitude for a star as one approaches the more heavily
crowded fields. 
 
\subsubsection{Comparison to the DENIS Catalog}

Finally, we compare our $I$-band photometry to that in the DENIS
catalog. Although the DENIS catalog is primarily an IR catalog, it
contains an $I$ band channel and \cite{c00} have extracted
point-source catalogs in the regions of the Magellanic Clouds.

\begin{figure}
\figurenum{17}
\plotone{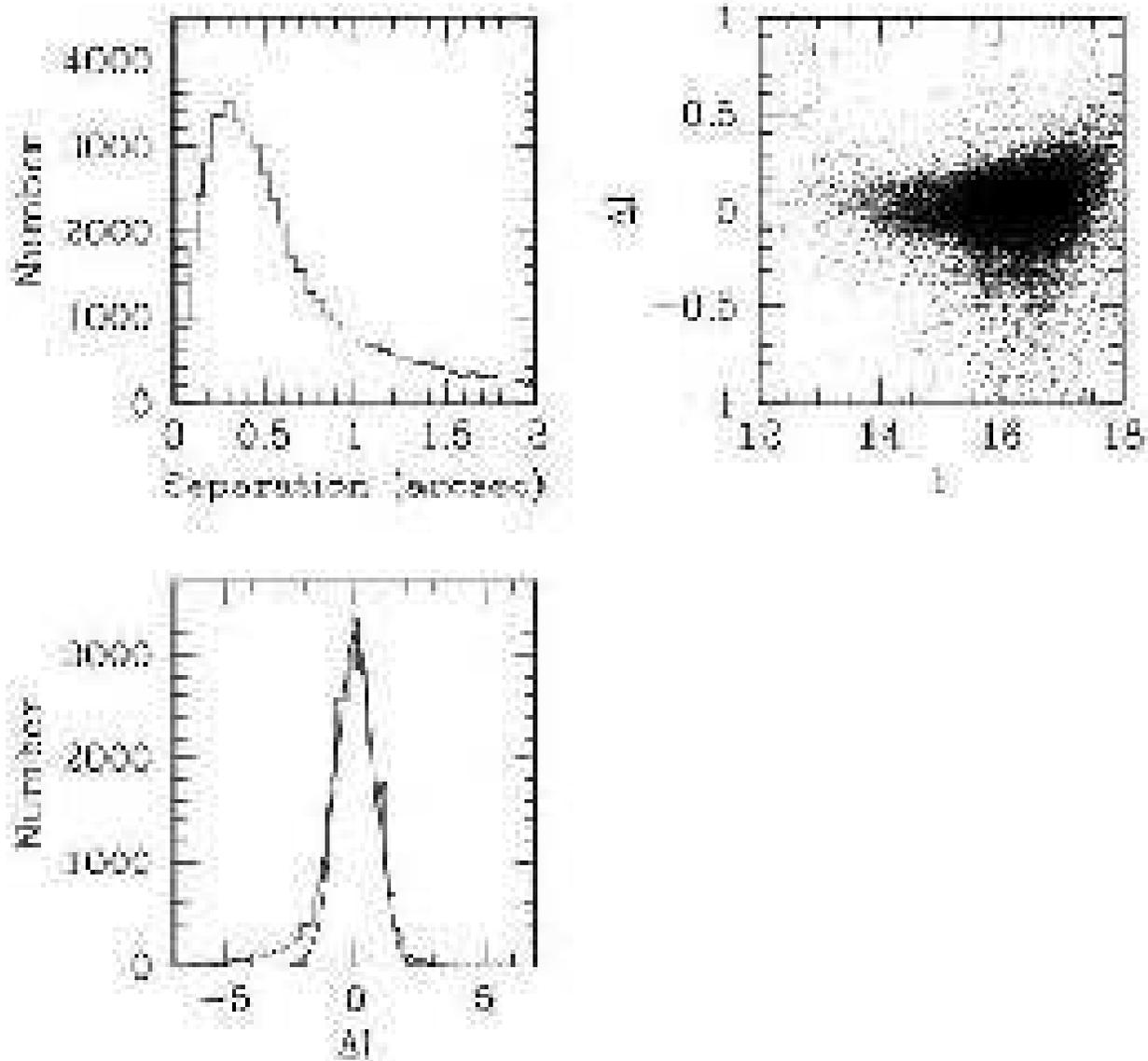}
\figcaption[deniscomp.ps]{The comparison of DENIS $I$ band catalog
with our data. The panel in the upper left illustrates the astrometric
differences for matched stars. The panel in the upper right
shows the difference in $I$ magnitudes as a function of $I$
magnitude. The panel in the lower left shows the differences
normalized by the calculated uncertainty. A Gaussian of unit
dispersion is overplotted for comparison.}
\label{deniscomp}
\end{figure}

We produce a similar comparison as to the Massey and OGLE
catalogs, using a search aperture of 3.5 arcsec for matches.
The distribution of astrometric and photometric differences for
matched stars are plotted in Figure~\ref{deniscomp}. In agreement with
our previous results, we find that the astrometric
accuracy is subpixel for the majority of the matches. The mean
difference is 0.9 arcsec, but the mode is $\sim 0.3$ arcsec. Using
only matched stars with positional offsets $<$ 1 pixel, the zeropoint
difference between the two surveys is $-0.008$ mag.  The distribution
of photometry differences, in units of standard deviations, is
entirely consistent with the propagated errors.

\section{Extinction Properties}

We developed a technique for fitting published stellar atmosphere
models (\cite{l97}) to $U, B, V$, and $I$ photometry to measure the
effective temperature, $T_E$, of the star and the line-of-sight extinction,
A$_V$ (\cite{z99}). We found that the model fitting was least
degenerate between $T_E$ and A$_V$ for stars with derived
temperatures in the ranges $5500 {\ \rm K}\le T_E \le 6500 {\ \ \rm K}$ and
$12,000 {\ \rm K}\le T_E \le 45000 {\ \rm K}$.  Therefore, 
we construct A$_V$ maps of the SMC
from the line-of-sight A$_V$ measurements to the set of ``cool''
stars ($5500 {\ \rm K}\le T_E \le 6500 {\ \rm K}$) and the set of ``hot''
stars ($12000 {\ \rm K}\le T_E \le 45000 {\ \rm K}$) with good quality
photometry ($\sigma_U < 0.4$, $\sigma_B < 0.2$, $\sigma_V < 0.2$,
$\sigma_I < 0.2$) and good model fits ($\chi^2 < 3$). In addition
to these criteria, we imposed a reddening-independent magnitude cut
($V < 19.0 + 3.2 \times (B-V)$).

\begin{figure}
\figurenum{18}
\vskip -2in
\plotone{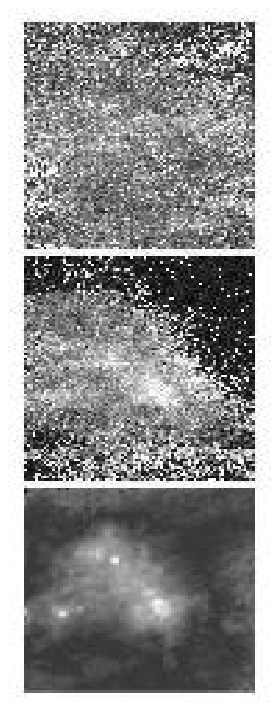}
\figcaption[extinctionmap.ps]{The spatial map of extinction
values as derived for both the cooler, older stars ($5500 {\ \rm K}\le T_E 
\le 6500 {\ \rm K}$ ; 
upper panel) and the hotter, younger stars ($12000 {\ \rm K}\le  T_E
\le 45000 {\ \rm K}$; middle panel) for the entire survey region. 
The scale on the upper panel spans $-0.1 \le A_V \le 0.4$, while the
middle panel spanse $-0.1 \le A_V \le 0.7$.
The localized circular regions of apparent high extinction are
globular clusters, which have some anomalous photometry because of their
high stellar densities. Horizontal variations are due to scan-to-scan 
photometry differences of a few hundreths of a magnitude. The upper left 
and bottom of the map for the hot population are dominated by a small
number of foreground stars. The lowest panel shows
the 100$\mu$m emission image from IRAS (not on exactly the same scale).}
\label{extinctionmap}
\end{figure}

The extinction maps
(Figure ~\ref{extinctionmap}) have lower signal-to-noise at large
projected radii from the SMC because there are fewer stars at those radii. In particular,
one should disregard the apparently low extinction toward the northwest
in the ``hot" population map because there are few such stars in this region.
Because the recovery of A$_V$
is quite sensitive to color, subtle differences in the scan
photometry are highlighted in the extinction maps (in particular
in that from the hot population). For example, a set of small photometric
differences (0.03 mag in opposite senses in $B$ and $I$, 
so that $B-I$ has changed by 0.06 mag) creates an extinction discontinuity
of the magnitude observed in the hot population map of Figure ~\ref{extinctionmap}.

The principal coherent extinction structure within the body of the SMC is the 
increase in extinction in the hot star population along the ridge of
the SMC (increasing to the southwest). This structure is not
visible
in the map from the colder stars indicating 1) that this structure may
not be real, 2) that most cold stars in the SMC are in the 
foreground relative to the hot stars, or 3) 
that the dust may be highly localized near the younger
stars, such that the lines-of-sight to background, colder stars are
unlikely to contain much dust.  We reject the first option
because this ridge corresponds well to the morphology of the 100$\mu$
IRAS emission (Figure~\ref{extinctionmap}, bottom panel). 
The second option appears
unlikely, simply because the colder, older stars are more likely to be
more extended along the line-of-sight, both toward the foreground and
background, than the younger stars. Therefore, we conclude that this observation
implies that the dust along the SMC ridge line is spatially highly localized around
the young, hot population. 

\begin{figure}
\figurenum{19}
\plotone{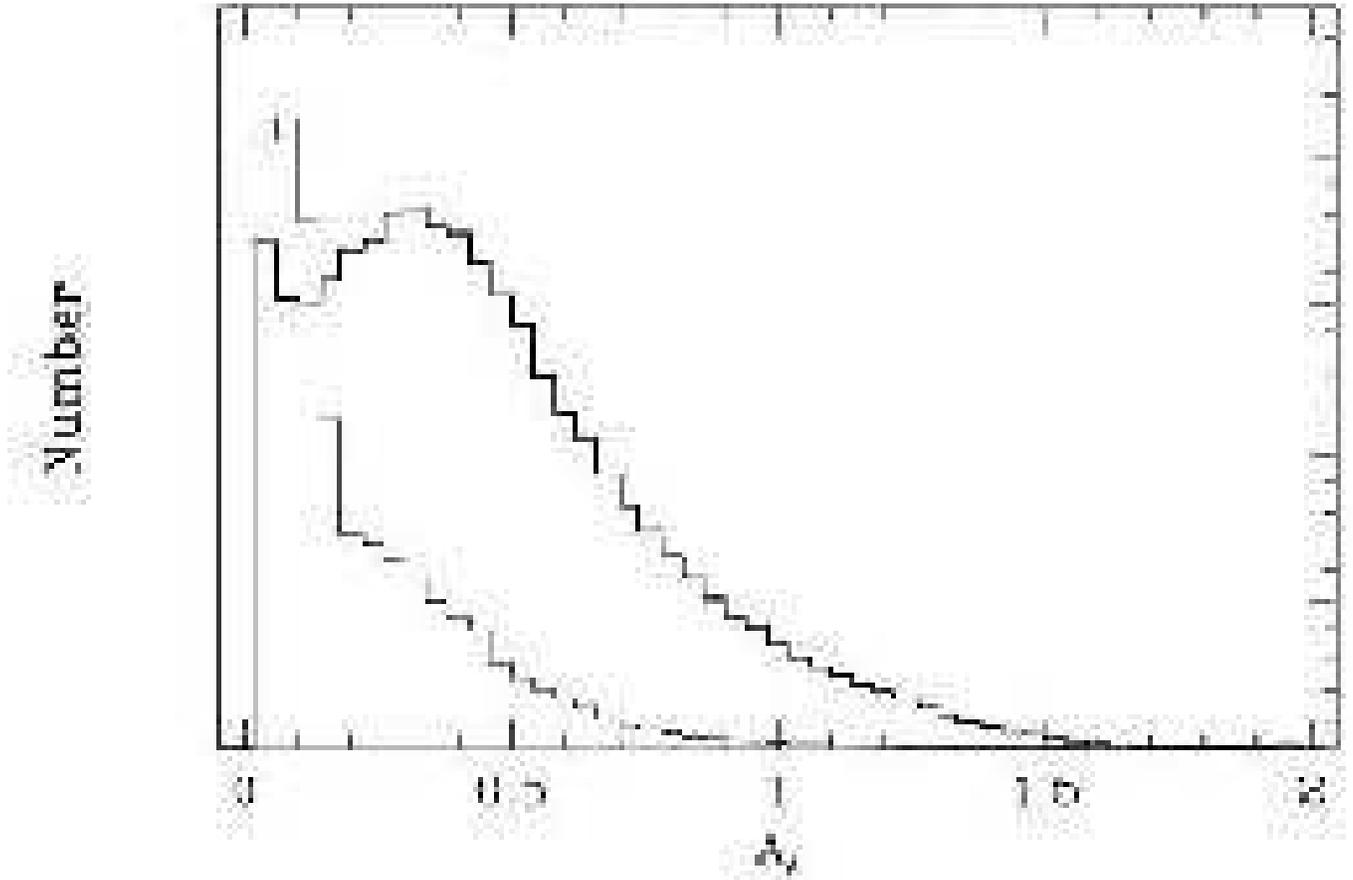}
\figcaption[extist.ps]{The distribution of extinction
values for the two stellar populations using stars with model
fits with $\chi^2 \le 3$ and the reddening-free magnitude
cut described in the text. 
The narrow line represents the distribution of 
$A_V$ for the 
colder population and the bolder line for the hotter population.
By construction $0 \le A_V < 2$. The two histograms are normalized relative to 
each other, but there are 319,705 stars in the colder population 
and 136,268 in the hotter. }
\label{exthist}
\end{figure}

The A$_V$ histograms of the two populations are shown in
Figure~\ref{exthist}. As we found for a region of the LMC (\cite{z99}),
the mean extinction is lower for the cooler
populations (average A$_V$ of 0.18 mag vs 0.46 mag for the cold vs. hot population,
respectively). Given a foreground extinction 
of $\sim$ 0.15 mag toward the SMC (\cite{bessel91}), the observed distribution
of extinctions toward the colder stars suggests that there is little, if any,
internal extinction in the SMC of the light from the
distributed older population.

The principal difference between the SMC and LMC A$_V$ histograms 
is the lack of a high-extinction tail for the cold stars in the SMC. 
For the LMC, this tail was interpreted to be the result of viewing roughly
half
of the colder stars through the irregular mid-plane dust layer associated with the 
hotter stars. We interpret the lack of such a tail in the SMC as an
indication that the SMC has no pervasive dust layer that extincts light
from background colder stars. This result is complementary to the 
explanation of the lack of a central ridge feature in the extinction map 
derived from the cold stars. 

The localization of the significant extinction near massive stars,
and the corresponding lack of significant extinction elsewhere in the
SMC, has several causes. The low value of the global extinction is in
part due to a dust-to-atomic gas ratio that is a factor of 30
below the Galactic value (\cite{s00}). The lower dust abundance is
related to the low observed abundance of CO in the SMC relative to the
Galaxy or LMC. The global scarcity of CO is thought to arise because
it is more easily photodissociated in the SMC (\cite{rlb93}). Lastly,
the CO that is present in the SMC is concentrated along the central
ridge, where most massive stars are found. We now have two cases (both
LMC and SMC) for which detailed line-of-sight extinction measurements
have shown significant differences ($\sim$ 0.3 mag) in the internal
extinction of young and old stars.  Typically, internal extinction
corrections for galaxies are derived from tracers of recent star
formation (for example, UV spectral slopes, $U-B$ colors, or H II
region line ratios).  Our results suggest that corrections derived
from such tracers are likely to be too large if applied to the entire
population of stars.

\section{Catalog Supplements}

The MCPS catalog for the SMC is presented as an ASCII text file (see
Table 1 for an example, the full catalog is provided electronically).
To enhance the usefulness of our catalog we augment the MCPS data in
several ways.  We have already described how we use data from
\cite{massey01} and \cite{udalski98} to correct our photometry for
the brightest 0.02\% of the cataloged stars ( $U, B, V$ for stars with
$B$ or $V < 13.5$ from Massey's catalog and $I$ for stars with $I <
13.5$ from OGLE).  Column 11 of the catalog (see Table 1) is a quality
flag that is the unique sum of several flags.  Stars for which we
replaced photometry with that from Massey's catalog have a quality
flag of $+1$. Stars for which we replaced photometry with that from
OGLE have a quality flag of $+2$ (a star with a quality flag $= 3$ has
had its original photometry replaced with that from both Massey's
catalog and OGLE).  Stars with measured magnitudes in $U, B, V$ and
$I$ and $\sigma_u < 0.4, \sigma_b < 0.2, \sigma_v < 0.2,$ and
$\sigma_I < 0.2$ are fit with stellar atmosphere models (\S4). Those
stars for which this fit is successful ($\chi^2 < 3$) are given a
quality flag $+10$ and those which are unsuccessful are given a
quality flag $+20$.  Therefore, a star with a quality flag of 11, for
example, had its $U, B$, and $V$ magnitudes replaced with those from
Massey's catalog and was successfully fit with a stellar atmosphere
model.

We supplement the MCPS with IR data from the DENIS and Two Micron All
Sky Survey (2MASS) surveys ($J$, $H$, and $K_S$, paired with
their uncertainties, are in columns 12 - 17).  We match stars in a
manner similar to our previous comparisons, except we use an
acceptance aperture of 1 arcsec and we have no magnitude criteria.  If
a star is matched only in 2MASS or DENIS data, then those data are
used. If a star is matched to both catalogs, we use the average
magnitude unless the IR magnitudes differed by more than 3$\sigma$
from each other in which case we do not use either data. We encourage
investigators to examine other combinations of the catalogs, for
example stars which are detected in the IR surveys but not in the
optical may provide an interesting sample of highly obscured stars.

\section{Summary}

We are conducting a broad-band photometric survey of the Magellanic
Clouds. Our intent is to provide these data to the community as quickly
as possible and here we present the data for over 5 million stars in
the $4.5^\circ \times 4^\circ$ survey area centered on the Small
Magellanic Cloud. To enable potential users to understand these data
we discuss various internal and external tests of the astrometry,
photometry, and associated uncertainties.  The catalog contains positions
(right ascension and declination in J2000 coordinates) and $U$, $B$, $V$, and $I$ magnitudes
and uncertainties in the Johnson-Kron-Cousins photometric system measured 
from our drift scan images. We augment this catalog with
indications of the data quality.  Lastly, we also provide
the 2MASS/DENIS IR magnitudes matched to our data for ease of use of a
dataset that is complete in $U$ to $K$ for 91,428 stars, provides
incomplete coverage through the infrared bands for 183,751 stars, and at 
least partial optical coverage for 5,156,057 stars.

Using this catalog, we have constructed extinction maps for two
stellar populations in the SMC.
We find 1) that the bulk of the dust is concentrated along the ridge of
the SMC, 2) that this dust is highly localized near the younger, hotter stars,
3) that aside from these regions of higher extinction, the average internal
extinction in the SMC is consistent with zero, and 4) that the average extinction
correction for the younger, hotter stars and the older, colder stars differs
by about 0.3 mag. The latter conclusion is  in agreement with what we found in the 
LMC (\cite{z99}). On a more general vein, the two external galaxies
for which we now have highly detailed maps of extinction as a function
of stellar population both show significant differences in the 
extinction toward those populations. This difference, or at least the
potential for this difference, should be considered when correcting
the photometry of other galaxies for internal extinction. Generalizing,
these results suggest that internal extinction corrections, 
which are commonly
derived from tracers of recent star formation, are typically overestimated.

\vskip 1in
\noindent
ACKNOWLEDGMENTS: DZ acknowledges financial support from an
NSF grant (AST-9619576), a NASA 
LTSA grant (NAG-5-3501), and fellowships from
the David and Lucile Packard Foundation and the Alfred P. Sloan
Foundation. EKG acknowledges support from NASA through 
grant HF-01108.01-98A from the Space Telescope Science Institute.

\clearpage

\clearpage

\clearpage

\figurenum{3}
\clearpage

\clearpage

\clearpage

\clearpage

\figurenum{7}
\clearpage

\clearpage

\clearpage

\clearpage

\clearpage

\clearpage

\clearpage

\clearpage

\clearpage

\clearpage

\clearpage

\clearpage

\clearpage

{\footnotesize{
\begin{deluxetable}{rrrrrrrrrrrrrrrrr}

\tablewidth{0pt}

\tablecaption{Sample of MCPS Catalog\tablenotemark{A}}

\tablehead{
\colhead{RA}&\colhead{Dec}&\colhead{$U$}&\colhead{$\sigma_U$}&\colhead{$B$}&\colhead{$\sigma_B$} &\colhead{$V$} & \colhead{$\sigma_V$}&\colhead{$I$}&\colhead{$\sigma_I$}&\colhead{Flag}&\colhead{$J$}&\colhead{$\sigma_J$}&\colhead{$H$}&\colhead{$\sigma_H$}&\colhead{$K_S$}&\colhead{$\sigma_{K_S}$} \\
}
\startdata 
0.398591&$-$74.09792&0.000&0.000&21.633&0.176&21.267&0.129&0.000&0.000&0&0.000&0.000&0.000&0.000&0.000&0.000\\
0.398591&$-$74.58766&0.000&0.000&22.394&0.165&22.446&0.223&0.000&0.000&0&0.000&0.000&0.000&0.000&0.000&0.000\\
0.398595&$-$73.92371&0.000&0.000&22.883&0.281&22.249&0.265&0.000&0.000&0&0.000&0.000&0.000&0.000&0.000&0.000\\
0.398595&$-$74.65561&19.804&0.201&20.845&0.095&20.715&0.075&20.625&0.160&10&0.000&0.000&0.000&0.000&0.000&0.000\\
0.398597&$-$74.01955&0.000&0.000&21.249&0.110&20.920&0.128&0.000&0.000&0&0.000&0.000&0.000&0.000&0.000&0.000\\
0.398598&$-$73.93850&0.000&0.000&21.362&0.144&21.120&0.133&0.000&0.000&0&0.000&0.000&0.000&0.000&0.000&0.000\\
0.398598&$-$74.32478&0.000&0.000&22.743&0.266&22.946&0.358&21.299&0.354&0&0.000&0.000&0.000&0.000&0.000&0.000\\
0.398600&$-$73.95302&17.380&0.058&17.819&0.033&17.927&0.030&17.996&0.042&10&0.000&0.000&0.000&0.000&0.000&0.000\\
0.398600&$-$74.01016&16.472&0.050&16.317&0.029&16.126&0.026&15.777&0.083&10&15.415&0.081&15.406&0.142&15.219&0.175\\
0.398603&$-$74.24788&0.000&0.000&22.945&0.278&21.286&0.080&19.681&0.080&0&0.000&0.000&0.000&0.000&0.000&0.000\\
0.398604&$-$74.37458&0.000&0.000&22.258&0.159&21.741&0.124&0.000&0.000&0&0.000&0.000&0.000&0.000&0.000&0.000\\
0.398604&$-$74.87499&0.000&0.000&22.628&0.214&21.801&0.119&21.172&0.230&0&0.000&0.000&0.000&0.000&0.000&0.000\\
0.398605&$-$74.74988&0.000&0.000&22.475&0.190&21.924&0.165&0.000&0.000&0&0.000&0.000&0.000&0.000&0.000&0.000\\
0.398606&$-$74.47800&18.868&0.071&18.837&0.058&18.059&0.023&17.108&0.048&10&16.350&0.111&15.792&0.164&0.000&9.999\\
0.398606&$-$74.78607&0.000&0.000&21.847&0.117&21.459&0.096&21.409&0.264&0&0.000&0.000&0.000&0.000&0.000&0.000\\
0.398607&$-$73.97942&0.000&0.000&21.651&0.112&20.748&0.091&19.818&0.089&0&0.000&0.000&0.000&0.000&0.000&0.000\\
\enddata
\tablenotetext{A}{The complete version of this table is in the electronic
edition of the Journal.  The printed edition contains only a sample.}
\end{deluxetable}
}}

\end{document}